\documentclass[aps,superscriptaddress,twocolumn,dvips]{revtex4}
\usepackage{feynmp}
\usepackage{amsmath}
\usepackage{amssymb}
\usepackage{epsfig}
\usepackage{graphicx}
\usepackage{subfigure}
\usepackage{hyperref}
\usepackage{bbm}
\usepackage{color}
\usepackage{longtable}
\usepackage{booktabs}
\usepackage{multirow}

\newcommand{\Rmnum}[1]{\expandafter\@slowromancap\romannumeral #1@}
\allowdisplaybreaks[3]
\begin{document}

\title{Possible instabilities in quadratic and cubic nodal line fermion systems with correlated interactions}

\author{Jing-Rong Wang}
\affiliation{Anhui Province Key Laboratory of Condensed Matter
Physics at Extreme Conditions, High Magnetic Field Laboratory of Anhui Province,
Chinese Academy of Sciences, Hefei 230031, China}
\author{Wei Li}
\altaffiliation{Corresponding author: wliustc@theory.issp.ac.cn}
\affiliation{Key Laboratory of Materials Physics,
Institute of Solid State Physics, Chinese Academy
of Sciences, Hefei 230031, China }
\author{Chang-Jin Zhang}
\altaffiliation{Corresponding author: zhangcj@hmfl.ac.cn}
\affiliation{Anhui Province Key Laboratory of Condensed Matter
Physics at Extreme Conditions, High Magnetic Field Laboratory of Anhui Province,
Chinese Academy of Sciences, Hefei 230031, China}
\affiliation{Institute of Physical Science and Information
Technology, Anhui University, Hefei 230601, China}

\begin{abstract}
Influence of  short-range four-fermion interactions on quadratic and cubic nodal line fermion
systems is studied by renormalization group theory. It is found that arbitrarily weak
four-fermion interaction could drive quadratic or  cubic nodal line fermion system
to a new phase. According to the initial conditions and value of fermion flavor, the system
may appear three kinds of instabilities.  First, quadratic or cubic nodal line is
split into conventional nodal lines, thus the system becomes nodal line semimetal.
Second, finite excitonic gap is generated, and the system becomes an excitonic insulator. Third, the
system is driven into superconducting phase. Thus, quadratic and cubic nodal
line fermion systems are rare strong correlated fermion systems in three dimension under the influence
of four-fermion interactions. These theoretical results may be verified in the candidates for quadratic
and cubic nodal line fermion systems.
\end{abstract}

\maketitle


\section{Introduction}

Study about topological semimetals (SMs) is one of the most important fields of modern
condensed matter physics \cite{Kotov12, Vafek14, Wehling14, Wan11, Weng16, FangChen16,
Yan17, Hasan17, Armitage18, WangJian18, LvQianDing19, Kruthoff17, Tang19Wan, Zhang19FangChen, VergnioryWangZ19}.
On the one hand, some SMs have wide potential industrial applications
due to their fantastic properties, such as large thermoelectric power \cite{Skinner18, Markov19}. On the other hand,
some SMs provide a platform to verify certain important concepts in high energy physics, due to that
their low-energy fermion excitations resemble the elementary particles \cite{Kotov12, Vafek14, Wehling14, Wan11, Weng16, FangChen16,
Yan17, Hasan17, Armitage18, WangJian18, LvQianDing19, Kruthoff17, Tang19Wan, Zhang19FangChen, VergnioryWangZ19}.

According to energy dispersion of the fermion excitations and topological property
of the system, SMs can be classified as Dirac SM (DSM), Weyl SM (WSM), multi-WSMs,
semi-DSM, Luttinger SM, nodal line SM (NLSM) \emph{etc}. \cite{Armitage18}. Graphene is a prototypical two dimensional (2D) DSM.
Angle-resolved photoemission spectroscopy (ARPES) experiments have confirmed that Cd$_{3}$As$_{2}$ and
Na$_{3}$Bi are 3D DSM \cite{Neupane14, Liu14}, and TaAs, TaP, NbAs, NbP are WSMs \cite{Hasan17, LvQianDing19, Xu15, Lv15}.
NLSM has been realized in PbTaSe$_{2}$, ZrSiS, ZrSiSe, HfSiS, and TiB$_{2}$
according to ARPES measurements \cite{Bian16, Neupane16, Schoop16, Hosen17, Takane16, Yi18, Liu18}.

In SMs, the dimension of Fermi surface is
at least two less than the dimension of the system. This characteristic is different
from conventional metals, in which the dimension of Fermi surface is one less than the dimension of the system \cite{GiulianiBook}.
In DSM, WSM, multi-WSMs, semi-DSM, and Luttinger SM,
the Fermi surface is composed by discrete points, where conduction and valence bands touch each other in the Brillouin zone. Whereas, the Fermi
surafce of NLSM is a line in the 3D Brillouin zone. Due to the abovementioned characteristic of
SMs, the density of states (DOS) of SMs  vanishes at the Fermi level.

Influence of interactions on SMs is an important and  nontrivial question, which attracts much attentions
\cite{Kotov12, Gonzalez99, Scheehy07, Hofmann14, WangLiu14, Goswami11, Hosur12, Throckmorton15, Moon13, Herbut14, YangNatPhys14,
Isobe16, WangLiuZhang17A, Lai15, Jian15, WangLiuZhang17B, ZhangShiXin17, WangLiuZhang18, Huh16, WangYuXuan17, WangLiuZhang19,
Han19, Zhang18, Han18A, Han18B, Roy18Birefringent, Uryszek19, Sur19, Herbut06, Herbut09, Maciejko14, Roy16, Sur16, Roy17A, Roy17B, Roy18A,  WangJing18, Roy18B}. Due to the vanishing
DOS, short-range four-fermion interaction is irrelevant in SM if it is weak, but may drive a quantum phase transition (QPT)
to a new phase if the interaction strength is large enough \cite{Herbut06, Herbut09, Maciejko14, Roy16, Sur16, Roy17A, Roy17B, Roy18A,  WangJing18, Roy18B}.
There have been studies on the effects of four-fermion interactions in
SMs including 2D DSM \cite{Herbut06, Herbut09}, 3D DSM \cite{Roy16}, WSM \cite{Maciejko14, Roy17B}, multi-WSMs \cite{Roy17B},
semi-DSM \cite{Roy18A, WangJing18}, Luttinger SM \cite{Roy18B}, and NLSM \cite{Sur16, Roy17A, Araujo18}. These studies showed that the SMs may be driven to
different phases according to the types of four-fermion interactions. Additionally, the influence of four-fermion interactions is closely
related to the fermion dispersion and topological property of the system.

In NLSM, the fermion dispersion is linear within the $x$-$y$ plane and also linear along the $z$ axis
\cite{Huh16, WangYuXuan17, Sur16, Roy17A, Araujo18, Syzranov17, LiXieGroup18, Chen19}.
For NLSM, DOS takes the form $\rho(\omega)\sim\omega$, which vanishes at the Fermi level, i.e.,  $\rho(0)=0$. Roy has analyzed
the possible QPTs in NLSM under the influence of four-fermion interactions \cite{Roy17A}.

Recently, Yu \emph{et al.} proposed that  quadratic and cubic nodal line fermion (NLF) systems could be realized in some materials \cite{Yu19}.
In these materials, the Fermi surface is also a line in the 3D Brillouin zone, which is same as NLSM. However, the fermion dispersion
is quadratic (cubic) within the $x$-$y$ plane and also quadratic (cubic) along the $z$ axis \cite{Yu19, LiLinHu17}. Accordingly, DOS satisfies $\rho(\omega)\sim\omega^{0}$ in quadratic
NLF system, and $\rho(\omega)\sim\omega^{-1/3}$ in cubic NLF system. Thus, the influence of four-fermion interactions on quadratic and
cubic NLF systems could be substantially different from the one in NLSM. This is an interesting and urgent question, which needs comprehensive
study.

In this article, we resolve this question through renormalization group (RG) theory \cite{Shankar94}. We find that quadratic and cubic
NLF fermion systems are unstable to short-range four-fermion interactions. We show that arbitrarily weak four-fermion interaction could drive
quadratic or cubic NLF system to a new phase. According to the initial conditions and value of fermion flavor, the system may appear three kinds of
instabilities. First, the quadratic or cubic nodal line is split into conventional nodal lines, and the system becomes a NLSM.
Second, finite excitonic gap is generated, then the system becomes an excitonic insulator. Third, the system is driven into a superconducting phase.

The rest of paper is structured as follows. The model is presented  in
Sec.~\ref{Sec:Model}. In Sec.~\ref{Sec:BillnearMeaning}, we analyze the physical meaning
of various fermion bilinears, which may be generated by the four-fermion interactions. We perform the mean field
calculation in Sec.~\ref{Sec:MeanFieldAnalysis}. In Sec.~\ref{Sec:Results}, we show RG
equations of the model parameters and  numerical results of the RG
equations. The behaviors of observable quantities in different phases are discussed in Sec.~\ref{Sec:ObserQuant}.
In Sec.~\ref{Sec:GeometryNL}, we discuss the role of geometry of nodal line.
The main results are summarized in Sec.~\ref{Sec:Summary}. The detailed calculation and
derivation for the RG equations are given in the Appendices.

\section{Model \label{Sec:Model}}

In Ref.~\cite{Yu19}, after performing a symmetry analysis over all 230 space groups for solid systems with spin-orbit coupling and
time-reversal symmetry, Yu \emph{et al.} found that quadratic and cubic nodal lines can be stabilized by crystalline symmetries.
For convenience, in Appendix~\ref{App:LatticeModel}, we show lattice models for quadratic and cubic NLF systems, and derive
the low-energy effective models. In the following, our analysis will focus on the low-energy
effective models directly.

The Hamiltonian density for free quadratic NLF system is given by
\begin{eqnarray}
\mathcal{H}_{0}^{q}(\mathbf{k})=A\left[\left(k_{r}^{2}-k_{z}^{2}\right)\sigma_{1}
+2k_{r}k_{z}\sigma_{2}\right], \label{Eq:HamiltonianDensityQNLF}
\end{eqnarray}
where $k_{r}=k_{\bot}-k_{F}$ with $k_{\bot}=\sqrt{k_{x}^{2}+k_{y}^{2}}$.
$A$ is a model parameter.
$\sigma_{1,2,3}$ are the standard Pauli matrices.
The energy spectrum for quadratic NLF is
\begin{eqnarray}
E_{q}(\mathbf{k})&=&\pm A\left(k_{r}^{2}+k_{z}^{2}\right). \label{Eq:DispersionQNLF}
\end{eqnarray}
For simplicity, here we do not consider
anisotropy of the fermion dispersion along $k_{r}$ and $k_{z}$. The qualitative
conclusions will not be changed if this anisotropy is incorporated.

The Hamiltonian density for cubic NLF system can be written as
\begin{eqnarray}
\mathcal{H}_{0}^{c}(\mathbf{k})=B\left[\left(k_{r}^{3}-3k_{r}k_{z}^{2}\right)\sigma_{1}
+\left(k_{z}^{3}-3k_{z}k_{r}^{2}\right)\sigma_{2}\right], \label{Eq:HamiltonianDensityCNLF}
\end{eqnarray}
with $B$ being a model parameter.
The energy dispersion for cubic NLF takes the form
\begin{eqnarray}
E_{c}(\mathbf{k})
&=&\pm B \left(k_{r}^{2}+k_{z}^{2}\right)^{3/2}.  \label{Eq:DispersionCNLF}
\end{eqnarray}

The Pauli matrices $\sigma_{i}$ act on the sublattice space of freedom.
Both the Hamiltonian densities $\mathcal{H}_{0}^{q}$ and  $\mathcal{H}_{0}^{c}$
satisfy the chiral symmetry $\left\{\mathcal{H}_{0}^{q,c},\sigma_{3}\right\}=0$. Once a term
$\mathcal{H}_{\Delta_{3}}=\Delta_{3}\sigma_{3}$ is generated, the fermions become gapped and
the chiral symmetry is broken \cite{Yu19}.

We consider the four-fermion interactions described by the action
\begin{eqnarray}
S_{\psi^{4}}&=&\frac{1}{N}\sum_{i=0}^{3}\lambda_{i}\int\frac{d\omega_{1}}{2\pi}\frac{d^{3}\mathbf{k}_{1}}{(2\pi)^{3}}
\frac{d\omega_{2}}{2\pi}\frac{d^{3}\mathbf{k}_{2}}{(2\pi)^{3}}
\frac{d\omega_{3}}{2\pi}\frac{d^{3}\mathbf{k}_{3}}{(2\pi)^{3}}\nonumber
\\
&&\times\psi^{\dag}(\omega_{1},\mathbf{k}_{1})\sigma_{i}
\psi(\omega_{2},\mathbf{k}_{2})\psi^{\dag}(\omega_{3},\mathbf{k}_{3})\sigma_{i}\nonumber
\\
&&\times\psi(\omega_{1}-\omega_{2}+\omega_{3},
\mathbf{k}_{1}-\mathbf{k}_{2}+\mathbf{k}_{3}),
\end{eqnarray}
where $\lambda_{i}$ with $i=0,1,2,3$ are the four-fermion coupling parameters, and $N$ is the
fermion flavor. The fermion flavor $N$ represents the degeneracy of the real spin. In the calculation, we take $N$ as a general turning
parameter. The physical value of $N$ is  $N=2$. $\sigma_{0}$ is the identity matrix. In the following, we are only interest in the case that
the initial value $\lambda_{i,0}$ satisfies $\lambda_{i,0}>0$, namely the interaction is repulsive initially.

\section{Physical meaning of fermion bilinears \label{Sec:BillnearMeaning}}

Decoupling  the four-fermion interactions, we could get four different fermion bilinears
$\psi^{\dag}\sigma_{0}\psi$, $\psi^{\dag}\sigma_{1}\psi$, $\psi^{\dag}\sigma_{2}\psi$, and $\psi^{\dag}\sigma_{3}\psi$.
The expectation values of these bilinears are given by
\begin{eqnarray}
\Delta_{0}&=&\left<\psi^{\dag}\sigma_{0}\psi\right>,
\\
\Delta_{1}&=&\left<\psi^{\dag}\sigma_{1}\psi\right>,
\\
\Delta_{2}&=&\left<\psi^{\dag}\sigma_{2}\psi\right>,
\\
\Delta_{3}&=&\left<\psi^{\dag}\sigma_{3}\psi\right>.
\end{eqnarray}
$\left<...\right>$ represents taking mean value on the ground state of total Hamiltonian.
They have different physical meanings. If $\Delta_{0}$ becomes finite, the Fermi level is modified, and
the Fermi surface is changed from 1D nodal line to 2D tube, since $\Delta_{0}$ represents the chemical potential.

The original nodal line is gapless for $(k_{r},k_{z})=(0,0)$. For quadratic NLF system, if  $\Delta_{1}$ becomes finite,
the original nodal line with quadratic dispersion is split into two
conventional nodal lines with linear dispersion. These two conventional nodal lines are gapless for the cases
\begin{eqnarray}
(k_{ar},k_{az})=\left(0,\left(\Delta_{1}/A\right)^{1/2}\right),
\end{eqnarray}
and
\begin{eqnarray}
(k_{br},k_{bz})=\left(0,-\left(\Delta_{1}/A\right)^{1/2}\right).
\end{eqnarray}
Around these two nodal lines, the fermion dispersion can be written as
\begin{eqnarray}
E&=&\pm\sqrt{4A\Delta_{1}\left(K_{r}^{2}+K_{z}^{2}\right)},
\end{eqnarray}
where  $K_{r}$ and $K_{z}$ are the momentum components relative to the nodal lines.

\begin{figure}[htbp]
\center
\includegraphics[width=3.1in]{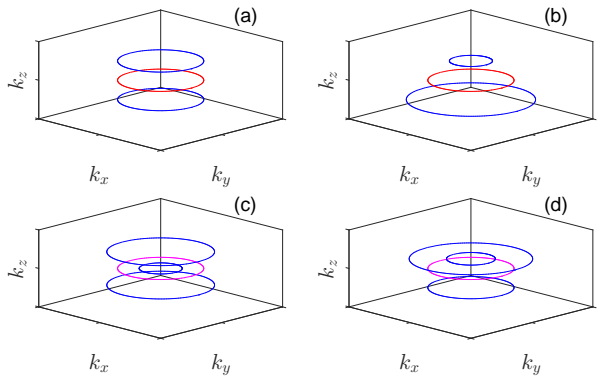}
\caption{ (a) and (b) Splitting of quadratic nodal line into conventional nodal lines in the presence
of $\Delta_{1}$ and $\Delta_{2}$ respectively. (c) and (d) Splitting of cubic nodal line into conventional nodal lines in the presence
of $\Delta_{1}$ and $\Delta_{2}$ respectively. Red, magenta, and blue lines correspond to
quadratic, cubic, and conventional nodal lines respectively.\label{Fig:SplitNodalLine}}
\end{figure}

For cubic NLF system, if $\Delta_{1}$ becomes finite, one cubic nodal line is split into three conventional nodal lines,
which are determined by
\begin{eqnarray}
\left(k_{ar},k_{az}\right)&=&\left(-\left(\Delta_{1}/B\right)^{1/3},0\right), \label{Eq:CubicNLFDelta1A}
\\
\left(k_{br},k_{bz}\right)&=&\left(\frac{1}{2}\left(\Delta_{1}/B\right)^{1/3}, \label{Eq:CubicNLFDelta1B}
\frac{\sqrt{3}}{2}\left(\Delta_{1}/B\right)^{1/3}\right),
\\
\left(k_{cr},k_{cr}\right)&=&\left(\frac{1}{2}\left(\Delta_{1}/B\right)^{1/3},
-\frac{\sqrt{3}}{2}\left(\Delta_{1}/B\right)^{1/3}\right). \label{Eq:CubicNLFDelta1C}
\end{eqnarray}
The energy dispersion around these three nodal lines can be expressed as
\begin{eqnarray}
E&=&\pm\sqrt{9B^{2/3}\Delta_{1}^{4/3}\left(K_{r}^{2}+K_{z}^{2}\right)}. \label{Eq:CubicNLFDelta1Dispersion}
\end{eqnarray}

For quadratic NLF system, if $\Delta_{2}$ acquires finite value, one quadratic nodal line is split into two conventional nodal lines,
which satisfy
\begin{eqnarray}
\left(k_{ar},k_{az}\right)&=&\left(\left(\Delta_{2}/(2A)\right)^{1/2},-\left(\Delta_{2}/(2A)\right)^{1/2}\right),
\\
\left(k_{br},k_{bz}\right)&=&\left(-\left(\Delta_{2}/(2A)\right)^{1/2},
\left(\Delta_{2}/(2A)\right)^{1/2}\right).
\end{eqnarray}
The energy dispersion of the fermions around these two nodal lines reads as
\begin{eqnarray}
E&=& \pm\sqrt{4A\Delta_{2}\left(K_{r}^{2}+K_{z}^{2}\right)}.
\end{eqnarray}

For cubic NLF system, if $\Delta_{2}$ becomes finite, one cubic nodal line is split into three conventional nodal lines.
The corresponding conditions can be obtained through Eq.~(\ref{Eq:CubicNLFDelta1A})-(\ref{Eq:CubicNLFDelta1C}) by
employing $k_{jz}\leftrightarrow k_{jr}$ with $j=a,b,c$ and $\Delta_{1}\rightarrow\Delta_{2}$.
Around these three nodal lines, the fermion dispersion can be expressed by   Eq.~(\ref{Eq:CubicNLFDelta1Dispersion}) through
utilizing $\Delta_{1}\rightarrow\Delta_{2}$.
.

If $\Delta_{3}$ acquires finite value, the energy dispersions for quadratic and cubic
NLF systems can be written as
\begin{eqnarray}
E_{q,c}(\mathbf{k},\Delta_{3})=\pm\sqrt{E_{q,c}^{2}(\mathbf{k})+\Delta_{3}^{2}},
\end{eqnarray}
where $E_{q}(\mathbf{k})$  and $E_{c}(\mathbf{k})$ are given by Eqs.~(\ref{Eq:DispersionQNLF}) and (\ref{Eq:DispersionCNLF})
respectively.
We can find that the fermion dispersion becomes gapped once $\Delta_{3}$ becomes
finite. Physically, it suggests that the system is driven into excitonic insulating phase.

For convenience, we show the splitting of quadratic and cubic nodal lines to conventional nodal lines in the presence of
$\Delta_{1}$ or $\Delta_{2}$ by the schematic diagrams in Fig.~\ref{Fig:SplitNodalLine}.

For a fermion system under the influence of four-fermion interaction $\lambda\left(\psi^{\dag}\Gamma\psi\right)^{2}$
where $\Gamma$ is a matrix, if the RG analysis shows that the four-fermion coupling strength $\lambda$ approaches
to infinity at a finite running parameter $\ell_{c}$, it indicates that the system becomes unstable under the energy scale
\begin{eqnarray}
\Lambda_{US}=\Lambda e^{-\ell_{c}}, \label{Eq:LambdaUS}
\end{eqnarray}
where $\Lambda$ ia an energy cutoff. For this situation, it is usually considered that a finite expectation value
$\Delta_{\Gamma}=\left<\psi^{\dag}\Gamma\psi\right>$ is generated. The magnitude of $\Delta_{\Gamma}$ can be
estimated through the energy scale $\Lambda_{US}$, i.e.,
\begin{eqnarray}
\Delta_{\Gamma}\sim\Lambda_{US}=\Lambda e^{-\ell_{c}}. \label{Eq:DeltaValue}
\end{eqnarray}
This method has been usually adopted in the RG studies about the
influence of four-fermion interactions on various fermion systems \cite{Herbut06, Herbut09, Maciejko14, Roy16, Sur16, Roy17A, Roy17B, Roy18A, WangJing18, Roy18B,
Nandkishore12, Metlitski15, Vafek10, Zhang10}.

If the four-fermion coupling parameter flows to negative infinity finally, we consider that the four-fermion interaction
becomes attractive in the low-energy regime. Accordingly, the system
is unstable to  pairing in the particle-particle channel, namely the generation of superconducting gap.

\section{Mean field analysis \label{Sec:MeanFieldAnalysis} }

Before performing the RG analysis, in this section, we analyze the generation of various order parameters under the influence of short-range four-fermion interactions
through mean field method .

\subsection{Quadratic NLF system}

\subsubsection{$\Delta_{1}=\left<\psi^{\dag}\sigma_{1}\psi\right>$ \label{Sec:QuadrticMean1}}

Considering  $\Delta_{1}$ induced by the four-fermion interaction $\lambda_{1}\left(\psi^{\dag}\sigma_{1}\psi\right)^{2}$, the fermion propagator in the Matsubara formalism
reads as
\begin{eqnarray}
G(\omega_{n},\mathbf{k})&=&\frac{1}{-i\omega_{n}+\mathcal{H}_{\mathbf{k},\Delta_{1}}^{q}},
\end{eqnarray}
where
\begin{eqnarray}
\mathcal{H}_{\mathbf{k},\Delta_{1}}^{q}=\left[A\left(k_{r}^{2}-k_{z}^{2}\right)+\Delta_{1}\right]\sigma_{1}
+2Ak_{r}k_{z}\sigma_{2}.
\end{eqnarray}
The mean field equation for $\Delta_{1}$ is given by
\begin{eqnarray}
\frac{\Delta_{1}}{\lambda_{1}}&=&T\sum_{\omega_{n}}\int\frac{d^3\mathbf{k}}{(2\pi)^{3}}\mathrm{Tr}\left[\sigma_{1}G(\omega_{n},\mathbf{k})\right]\nonumber
\\
&=&2T\int\frac{d^3\mathbf{k}}{(2\pi)^{3}}\left[A\left(k_{r}^{2}-k_{z}^{2}\right)+\Delta_{1}\right]\nonumber
\\
&&\times\sum_{\omega_{n}}\frac{1}
{\omega_{n}^{2}+E_{\mathbf{k},\Delta_{1}}^{2}},
\end{eqnarray}
where $\omega_{n}=(2n+1)\pi T$ with $n$ being integers, and
\begin{eqnarray}
E_{\mathbf{k},\Delta_{1}}=\sqrt{A^{2}\left(k_{r}^{2}+k_{z}^{2}\right)^{2}+2A\left(k_{r}^{2}-k_{z}^{2}\right)\Delta_{1}+\Delta_{1}^{2}}.
\end{eqnarray}
Performing the frequency summation, we get
\begin{eqnarray}
\Delta_{1}&=&\int\frac{d^3\mathbf{k}}{(2\pi)^{3}}\left[A\left(k_{r}^{2}-k_{z}^{2}\right)+\Delta_{1}\right]
\frac{1}{E_{\mathbf{k},\Delta_{1}}}\nonumber
\\
&&\times\tanh\left( \frac{E_{\mathbf{k},\Delta_{1}}}{2T}\right).
\end{eqnarray}
Linearizing $\Delta_{1}$ in the vicinity of critical temperature $T_{c}$ for the phase transition, we obtain the equation for $T_{c}$
\begin{eqnarray}
\frac{\Delta_{1}}{\lambda_{1}}
&=&\Delta_{1}\int\frac{d^3\mathbf{k}}{(2\pi)^{3}}
\frac{1}{E_{q}(\mathbf{k})}\nonumber
\\
&&\times\Bigg\{\left[1-\frac{\left(k_{r}^{2}-k_{z}^{2}\right)^{2}}{\left(k_{r}^{2}+k_{z}^{2}\right)^{2}}\right]
\tanh\left( \frac{E_{q}(\mathbf{k})}{2T_{c}}\right)\nonumber
\\
&&+\frac{E_{q}(\mathbf{k})}{2T_{c}}
\frac{1}{\cosh^{2}\left( \frac{E_{q}(\mathbf{k})}{2T_{c}}\right)}\frac{\left(k_{r}^{2}-k_{z}^{2}\right)^{2}}
{\left(k_{r}^{2}+k_{z}^{2}\right)^{2}}\Bigg\}.
\end{eqnarray}
The equation can be further written as
\begin{eqnarray}
\frac{1}{\lambda_{1}}&=&\frac{k_{F}}{4\pi}\int dKK
\frac{1}{AK^{2}}
\Bigg[
\tanh\left( \frac{AK^{2}}{2T_{c}}\right)\nonumber
\\
&&+ \frac{AK^{2}}{2T_{c}}
\frac{1}{\cosh^{2}\left( \frac{AK^{2}}{2T_{c}}\right)}\Bigg]\nonumber
\\
&=&\frac{k_{F}}{8\pi A}\Bigg[\ln\left(\frac{A\Lambda^{2}}{2T_{c}}\right)
\tanh\left(\frac{A\Lambda^{2}}{2T_{c}}\right)-\int_{0}^{\frac{A\Lambda^{2}}{2T_{c}}} dx\ln(x)\nonumber
\\
&&\times\frac{1}{\cosh^{2}(x)}+\tanh\left(\frac{A\Lambda^{2}}{2T_{c}}\right)\Bigg].
\end{eqnarray}
If $T_{c}\ll A\Lambda^{2}$, the equation becomes
\begin{eqnarray}
\frac{1}{\lambda_{1}}
&\approx&\frac{k_{F}}{8\pi A}\left[\ln\left(\frac{A\Lambda^{2}}{2T_{c}}\right)-\int_{0}^{+\infty} dx\ln(x)\frac{1}{\cosh^{2}(x)}+1\right]\nonumber
\\
&=&\frac{k_{F}}{8\pi A}\ln\left[\left(\frac{A\Lambda^{2}}{T_{c}}\right)\left(\frac{2e^{\gamma+1}}{\pi}\right)\right],
\end{eqnarray}
where $\gamma$ represents the Euler constant, and it satisfies $\gamma\approx0.577215$.
Thus $T_{c}$ is given by
\begin{eqnarray}
T_{c}=\frac{2e^{\gamma+1}}{\pi}A\Lambda^{2}e^{-\frac{k_{F}}{8\lambda_{1}\pi A}}.
\end{eqnarray}

\subsubsection{$\Delta_{2}=\left<\psi^{\dag}\sigma_{2}\psi\right>$}

Incorporating $\Delta_{2}$ induced by the four-fermion interaction $\lambda_{2}\left(\psi^{\dag}\sigma_{2}\psi\right)^{2}$, after similar derivation
shown in subsection~\ref{Sec:QuadrticMean1},
we obtain the critical temperature for the phase transition as
\begin{eqnarray}
T_{c}=\frac{2e^{\gamma+1}}{\pi}A\Lambda^{2}e^{-\frac{k_{F}}{8\lambda_{2}\pi A}}.
\end{eqnarray}

\subsubsection{$\Delta_{3}=\left<\psi^{\dag}\sigma_{3}\psi\right>$}

 Considering $\Delta_{3}$ induced by the four-fermion interaction $\lambda_{3}\left(\psi^{\dag}\sigma_{3}\psi\right)^{2}$, the fermion propagator
 in the Matsubara formalism takes the form
\begin{eqnarray}
G(\omega_{n},\mathbf{k})=\frac{1}{-i\omega_{n}+\mathcal{H}_{\mathbf{k},\Delta_{3}}^{q}},
\end{eqnarray}
where
\begin{eqnarray}
\mathcal{H}_{\mathbf{k},\Delta_{3}}^{q}&=&A\left[\left(k_{r}^{2}-k_{z}^{2}\right)\sigma_{1}
+2k_{r}k_{z}\sigma_{2}\right]+\Delta_{3}\sigma_{3}.
\end{eqnarray}
The mean field equation for $\Delta_{3}$ is determined by
\begin{eqnarray}
\frac{\Delta_{3}}{\lambda_{3}}&=&T\sum_{\omega_{n}}\int\frac{d^3\mathbf{k}}{(2\pi)^{3}}\mathrm{Tr}\left[\sigma_{3}G(\omega_{n},\mathbf{k})\right]\nonumber
\\
&=&2\Delta_{3}T\int\frac{d^3\mathbf{k}}{(2\pi)^{3}}\sum_{\omega_{n}}\frac{1}{\omega_{n}^{2}+E_{\mathbf{k},\Delta_{3}}^{2}},
\end{eqnarray}
where
\begin{eqnarray}
E_{\mathbf{k},\Delta_{3}}=\sqrt{A^{2}\left(k_{r}^{2}+k_{z}^{2}\right)^{2}+\Delta_{3}^{2}}.
\end{eqnarray}
Carrying out the frequency summation, we obtain
\begin{eqnarray}
\frac{\Delta_{3}}{\lambda_{3}}
&=&\Delta_{3}\int\frac{d^3\mathbf{k}}{(2\pi)^{3}}\frac{1}{E_{\mathbf{k},\Delta_{3}}}
\tanh\left( \frac{E_{\mathbf{k},\Delta_{3}}}{2T}\right),
\end{eqnarray}
which can be further written as
\begin{eqnarray}
\frac{1}{\lambda_{3}}
&=&\frac{k_{F}}{2\pi}\int_{0}^{\Lambda} dKK\frac{1}{\sqrt{A^{2}K^{4}+\Delta_{3}^{2}}}\nonumber
\\
&&\times\tanh\left( \frac{\sqrt{A^{2}K^{4}+\Delta_{3}^{2}}}{2T}\right).
\end{eqnarray}
Linearizing $\Delta_{3}$ in the vicinity of $T_{c}$ yields
\begin{eqnarray}
\frac{1}{\lambda_{3}}
&=&\frac{k_{F}}{2\pi}\int_{0}^{\Lambda} dKK\frac{1}{AK^{2}}
\tanh\left( \frac{AK^{2}}{2T_{c}}\right)\nonumber
\\
&=&\frac{k_{F}}{4\pi A}\Bigg[\ln\left(\frac{A\Lambda^{2}}{2T_{c}}\right)
\tanh\left(\frac{A\Lambda^{2}}{2T_{c}}\right)\nonumber
\\
&&-\int_{0}^{\frac{A\Lambda^{2}}{2T_{c}}} dx\ln(x)\frac{1}{\cosh^{2}(x)}\Bigg].
\end{eqnarray}
If $T_{c}\ll A\Lambda^{2}$, we have
\begin{eqnarray}
\frac{1}{\lambda_{3}}
&\approx&\frac{k_{F}}{4\pi A}\left[\ln\left(\frac{A\Lambda^{2}}{2T_{c}}\right)-\int_{0}^{+\infty} dx\ln(x)\frac{1}{\cosh^{2}(x)}\right]\nonumber
\\
&=&\frac{k_{F}}{4\pi A}\ln\left[\left(\frac{A\Lambda^{2}}{T_{c}}\right)\left(\frac{2e^{\gamma}}{\pi}\right)\right].
\end{eqnarray}
Namely
\begin{eqnarray}
T_{c}=\frac{2e^{\gamma}}{\pi}A\Lambda^{2}e^{-\frac{k_{F}}{4\lambda_{3}\pi A}}.
\end{eqnarray}

\subsection{Cubic NLF system}

\subsubsection{$\Delta_{1}=\left<\psi^{\dag}\sigma_{1}\psi\right>$ \label{Sec:CubicMean1}}

Considering $\Delta_{1}$ generated by the  four-fermion interaction $\lambda_{1}\left(\psi^{\dag}\sigma_{1}\psi\right)^{2}$,
the fermion propagator in Matsubara formalism can be written as
\begin{eqnarray}
G(\omega_{n},\mathbf{k})=\frac{1}{-i\omega_{n}+\mathcal{H}_{\mathbf{k},\Delta_{1}}^{c}},
\end{eqnarray}
where
\begin{eqnarray}
\mathcal{H}_{\mathbf{k},\Delta_{1}}^{c}&=&\left[B\left(k_{r}^{3}-3k_{r}k_{z}^{2}\right)+\Delta_{1}\right]\sigma_{1}\nonumber
\\
&&+B\left(k_{z}^{3}-3k_{z}k_{r}^{2}\right)\sigma_{2}.
\end{eqnarray}
The mean field equation for $\Delta_{1}$ takes the form
\begin{eqnarray}
\frac{\Delta_{1}}{\lambda_{1}}&=&T\sum_{\omega_{n}}\int\frac{d^3\mathbf{k}}{(2\pi)^{3}}\mathrm{Tr}
\left[\sigma_{1}G(\omega_{n},\mathbf{k})\right]\nonumber
\\
&=&2T\int\frac{d^3\mathbf{k}}{(2\pi)^{3}}\left[B\left(k_{r}^{3}-3k_{r}k_{z}^{2}\right)+\Delta_{1}\right]\nonumber
\\
&&\times\sum_{\omega_{n}}\frac{1}{\omega_{n}^{2}
+E_{\mathbf{k},\Delta_{1}}^{2}},
\end{eqnarray}
where
\begin{eqnarray}
E_{\mathbf{k},\Delta_{1}}&=&\sqrt{B^{2}\left(k_{r}^{2}+k_{z}^{2}\right)^{3}+2B\left(k_{r}^{3}-3k_{r}k_{z}^{2}\right)\Delta_{1}+\Delta_{1}^{2}}.\nonumber
\\
\end{eqnarray}
Carrying out the frequency summation gives to
\begin{eqnarray}
\frac{\Delta_{1}}{\lambda_{1}}
&=&\int\frac{d^3\mathbf{k}}{(2\pi)^{3}}\left[B\left(k_{r}^{3}-3k_{r}k_{z}^{2}\right)+\Delta_{1}\right]
\frac{1}{E_{\mathbf{k},\Delta_{1}}}\nonumber
\\
&&\times\tanh\left( \frac{E_{\mathbf{k},\Delta_{1}}}{2T}\right).
\end{eqnarray}
Linearizing $\Delta_{1}$ in the vicinity of $T_{c}$, we find
\begin{eqnarray}
\frac{\Delta_{1}}{\lambda_{1}}
&=&\Delta_{1}\int\frac{d^3\mathbf{k}}{(2\pi)^{3}}
\frac{1}{E_{c}(\mathbf{k})}\nonumber
\\
&&\times\Bigg\{ \left[1-\frac{\left(k_{r}^{3}-3k_{r}k_{z}^{2}\right)^{2}}{\left(k_{r}^{2}+k_{z}^{2}\right)^{3}}\right]
\tanh\left(\frac{E_{c}(\mathbf{k})}{2T_{c}}\right)\nonumber
\\
&&+\frac{E_{c}(\mathbf{k})}{2T_{c}}
\frac{1}{\cosh^{2}\left(\frac{E_{c}(\mathbf{k})}{2T_{c}}\right)}
\frac{\left(k_{r}^{3}-3k_{r}k_{z}^{2}\right)^{2}
}{\left(k_{r}^{2}+k_{z}^{2}\right)^{3}}\Bigg\}.
\end{eqnarray}
This equation can be further expressed as
\begin{eqnarray}
\frac{1}{\lambda_{1}}
&=&\frac{k_{F}}{4\pi}\int_{0}^{\Lambda} dKK\frac{1}{BK^{3}}
\tanh\left( \frac{BK^{3}}{2T_{c}}\right)\nonumber
\\
&=&\frac{k_{F}}{12\pi B^{2/3}(2T_{c})^{1/3}}\Bigg[(-3)\frac{1}{\left(\frac{B\Lambda^{3}}{2T_{c}}\right)^{1/3}}
\tanh\left(\frac{B\Lambda^{3}}{2T_{c}}\right)\nonumber
\\
&&+4\int_{0}^{\frac{B\Lambda^{3}}{2T_{c}}}dx\frac{1}{x^{1/3}}
\frac{1}{\cosh^{2}\left(x\right)}\Bigg].
\end{eqnarray}
In the limit $T_{c}\ll B\Lambda^{3}$, we get
\begin{eqnarray}
\frac{1}{\lambda_{1}}
&\approx&\frac{k_{F}}{12\pi B^{2/3}(2T_{c})^{1/3}}4\int_{0}^{+\infty}dx\frac{1}{x^{1/3}}
\frac{1}{\cosh^{2}\left(x\right)}\nonumber
\\
&=&\frac{ak_{F}}{3\cdot2^{1/3}\pi B^{2/3}T_{c}^{1/3}},
\end{eqnarray}
where
\begin{eqnarray}
a=\int_{0}^{+\infty}dx\frac{1}{x^{1/3}}
\frac{1}{\cosh^{2}\left(x\right)}\approx1.43829.
\end{eqnarray}
Accordingly, $T_{c}$ can be expressed as
\begin{eqnarray}
T_{c}
&=&\left(\frac{a\lambda_{1}k_{F}}{3\cdot2^{1/3}\pi B^{2/3}}\right)^{3}.
\end{eqnarray}

\subsubsection{$\Delta_{2}=\left<\psi^{\dag}\sigma_{2}\psi\right>$}

Considering $\Delta_{2}$ induced by the four-fermion interaction $\lambda_{2}\left(\psi^{\dag}\sigma_{2}\psi\right)^{2}$,
similar to subsection~\ref{Sec:CubicMean1},
we obtain
\begin{eqnarray}
T_{c}
&=&\left(\frac{a\lambda_{2}k_{F}}{3\cdot2^{1/3}\pi B^{2/3}}\right)^{3}.
\end{eqnarray}

\subsubsection{$\Delta_{3}=\left<\psi^{\dag}\sigma_{3}\psi\right>$}

Incorporating $\Delta_{3}$ induced by the four-fermion interaction $\lambda_{3}\left(\psi^{\dag}\sigma_{3}\psi\right)^{2}$, the fermion propagator in
the Matsubara formalism reads as
\begin{eqnarray}
G(\omega_{n},\mathbf{k})=\frac{1}{-i\omega_{n}+\mathcal{H}_{\mathbf{k},\Delta_{3}}^{c}
},
\end{eqnarray}
where
\begin{eqnarray}
\mathcal{H}_{\mathbf{k},\Delta_{3}}^{c}&=&B\left[\left(k_{r}^{3}-3k_{r}k_{z}^{2}\right)\sigma_{1}
+\left(k_{z}^{3}-3k_{z}k_{r}^{2}\right)\sigma_{2}\right]\nonumber
\\
&&+\Delta_{3}\sigma_{3}.
\end{eqnarray}
The mean field equation for $\Delta_{3}$ is given by
\begin{eqnarray}
\frac{\Delta_{3}}{\lambda_{3}}&=&T\sum_{\omega_{n}}\int\frac{d^3\mathbf{k}}{(2\pi)^{3}}\mathrm{Tr}
\left[\sigma_{3}G_{0}(\omega_{n},\mathbf{k})\right]\nonumber
\\
&=&2\Delta_{3}T\int\frac{d^3\mathbf{k}}{(2\pi)^{3}}\sum_{\omega_{n}}\frac{1}{\omega_{n}^{2}
+E_{\mathbf{k},\Delta_{3}}^{2}},
\end{eqnarray}
where
\begin{eqnarray}
E_{\mathbf{k},\Delta_{3}}=\sqrt{B^{2}\left(k_{r}^{2}+k_{z}^{2}\right)^{3}+\Delta_{3}^{2}}.
\end{eqnarray}
Performing the frequency summation, we get
\begin{eqnarray}
\frac{\Delta_{3}}{\lambda_{3}}
&=&\Delta_{3}\int\frac{d^3\mathbf{k}}{(2\pi)^{3}}\frac{1}{E_{\mathbf{k},\Delta_{3}}}
\tanh\left( \frac{E_{\mathbf{k},\Delta_{3}}}{2T}\right).
\end{eqnarray}
The equation can be further written as
\begin{eqnarray}
\frac{1}{\lambda_{3}}
&=&\frac{k_{F}}{2\pi}\int_{0}^{\Lambda} dKK\frac{1}{\sqrt{B^{2}K^{6}+\Delta_{3}^{2}}}\nonumber
\\
&&\times\tanh\left( \frac{\sqrt{B^{2}K^{6}+\Delta_{3}^{2}}}{2T}\right).
\end{eqnarray}
$T_{c}$ is determined by
\begin{eqnarray}
\frac{1}{\lambda_{3}}
&=&\frac{k_{F}}{2\pi}\int_{0}^{\Lambda} dKK\frac{1}{BK^{3}}
\tanh\left( \frac{BK^{3}}{2T_{c}}\right)\nonumber
\\
&=&\frac{k_{F}}{6\pi B^{2/3}(2T_{c})^{1/3}}\Bigg[(-3)\frac{1}{\left(\frac{B\Lambda^{3}}{2T_{c}}\right)^{1/3}}
\tanh\left(\frac{B\Lambda^{3}}{2T_{c}}\right)\nonumber
\\
&&+3\int_{0}^{\frac{B\Lambda^{3}}{2T_{c}}}dx\frac{1}{x^{1/3}}
\frac{1}{\cosh^{2}\left(x\right)}\Bigg].
\end{eqnarray}
In the limit $T_{c}\ll B\Lambda^{3}$, we obtain
\begin{eqnarray}
\frac{1}{\lambda_{3}}
&\approx&\frac{k_{F}}{6\pi B^{2/3}(2T_{c})^{1/3}}3\int_{0}^{+\infty}dx\frac{1}{x^{1/3}}
\frac{1}{\cosh^{2}\left(x\right)}\nonumber
\\
&=&\frac{ak_{F}}{2^{4/3}\pi B^{2/3}T_{c}^{1/3}},
\end{eqnarray}
which is equivalent to
\begin{eqnarray}
T_{c}
&=&\left(\frac{a\lambda_{3}k_{F}}{2^{4/3}\pi B^{2/3}}\right)^{3}.
\end{eqnarray}

\begin{figure}[htbp]
\center
\includegraphics[width=3.1in]{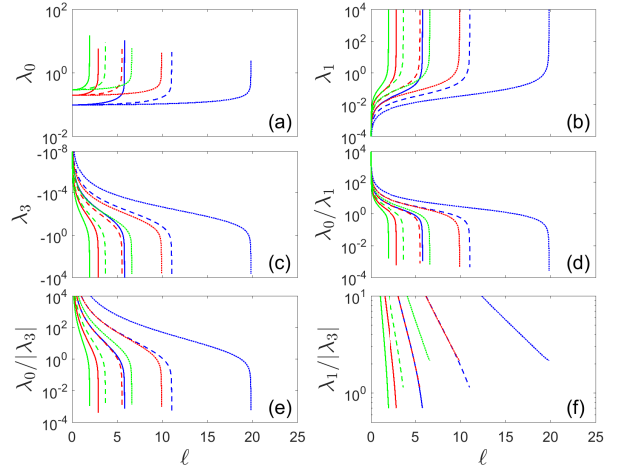}
\caption{Flows of $\lambda_{0}$, $\lambda_{1}$, $\lambda_{3}$, $\lambda_{0}/\lambda_{1}$, $\lambda_{0}/|\lambda_{3}|$,
and $\lambda_{1}/|\lambda_{3}|$ in quadratic NLF system. Blue, red, and green colors correspond to $\lambda_{0,0}=0.1, 0.2, 0.3$ respectively.
Solid, dashed, and dotted lines stand for $N=1, 2, 4$ respectively. \label{Fig:VRGQNLFInilambda0}}
\end{figure}

\section{Renormalization group analysis \label{Sec:Results}}

In this section, we present the RG results of the influence of
four-fermion interactions on quadratic and cubic NLF systems.
The detailed derivation for the RG equations are shown in the Appendices.

\subsection{Quadratic NLF}

For quadratic NLF system in the presence of four-fermion interactions, the RG equations for the coupling  parameters
are given by
\begin{eqnarray}
\frac{d\lambda_{0}}{d\ell}&=&
\left(\lambda_{0}\lambda_{1}+\lambda_{0}\lambda_{2}\right)\frac{1}{N}, \label{Eq:RGElambda0QNLSM}
\\
\frac{d\lambda_{1}}{d\ell}&=& \bigg[\left(\lambda_{0}^{2}+\lambda_{2}^{2}
+\lambda_{3}^{2}+
\lambda_{0}\lambda_{1}+2\lambda_{1}\lambda_{2}+\lambda_{1}\lambda_{3}\right.\nonumber
\\
&&\left.-2\lambda_{2}\lambda_{3}
\right)\frac{1}{N}+\lambda_{1}^{2}\bigg], \label{Eq:RGElambda1QNLSM}
\\
\frac{d\lambda_{2}}{d\ell}&=&\bigg[\left(\lambda_{0}^{2}+\lambda_{1}^{2}
+\lambda_{3}^{2}+\lambda_{0}\lambda_{2}+2\lambda_{1}\lambda_{2}-2\lambda_{1}\lambda_{3}\right.\nonumber
\\
&&\left.
+\lambda_{2}\lambda_{3}\right)
\frac{1}{N}+\lambda_{2}^{2}\bigg], \label{Eq:RGElambda2QNLSM}
\\
\frac{d\lambda_{3}}{d\ell}&=&\bigg[\left(-2\lambda_{3}^{2}+2\lambda_{0}\lambda_{3}-2\lambda_{1}\lambda_{2}+3\lambda_{1}\lambda_{3}+3\lambda_{2}\lambda_{3}
\right)\nonumber
\\
&&\times\frac{1}{N}+2\lambda_{3}^{2}\bigg]. \label{Eq:RGElambda3QNLSM}
\end{eqnarray}
The transformations $\frac{k_{F}}{2\pi A}\lambda_{i}\rightarrow\lambda_{i}$ with $i=0,1,2,3$
have been employed in the derivation of the RG equations. We notice that $\lambda_{0}$ is not
generated if the initial value $\lambda_{0,0}=0$.

\begin{figure}[htbp]
\center
\includegraphics[width=3.1in]{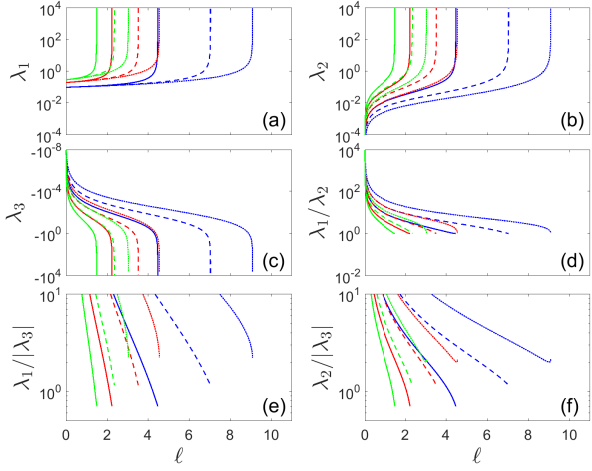}
\caption{Flows of $\lambda_{1}$, $\lambda_{2}$, $\lambda_{3}$, $\lambda_{1}/\lambda_{2}$, $\lambda_{1}/|\lambda_{3}|$,
and $\lambda_{2}/|\lambda_{3}|$ in quadratic NLF system. Blue, red, and green colors correspond to $\lambda_{1,0}=0.1, 0.2, 0.3$ respectively.
Solid, dashed, and dotted lines stand for $N=1, 2, 4$ respectively. \label{Fig:VRGQNLFInilambda1}}
\end{figure}

If the initial value $\lambda_{0,0}$ is finite, the flows of $\lambda_{0}$, $\lambda_{1}$, $\lambda_{3}$, $\lambda_{0}/\lambda_{1}$,
$\lambda_{0}/|\lambda_{3}|$, $\lambda_{1}/|\lambda_{3}|$ are shown in Fig.~\ref{Fig:VRGQNLFInilambda0}.
As shown in Figs.~\ref{Fig:VRGQNLFInilambda0}(a)-\ref{Fig:VRGQNLFInilambda0}(c), $\lambda_{0}$, $\lambda_{1}$ approach to infinity, and $\lambda_{3}$ flows to negative infinity
at same finite energy scale. According to Figs.~\ref{Fig:VRGQNLFInilambda0}(d)-\ref{Fig:VRGQNLFInilambda0}(f), $\lambda_{0}/\lambda_{1}$ and
$\lambda_{0}/|\lambda_{3}|$ approach to zero. Additionally, $\lambda_{1}/|\lambda_{3}|$ flows to a constant smaller than $1$ for $N=1$, but
approaches to a constant larger than 1 for $N\ge2$. $\lambda_{1}/\lambda_{2}$ is always equal to 1, which is not shown in Fig.~\ref{Fig:VRGQNLFInilambda0}.
These results indicate that  arbitrarily weak
four-fermion interaction induces the system to be unstable. For $N=1$, generation of
superconducting gap is the leading instability. However, splitting of quadratic nodal line with generation of $\Delta_{1}$ or $\Delta_{2}$ is
the subleading instability. For $N\ge2$, splitting of quadratic nodal line with generation of $\Delta_{1}$ or $\Delta_{2}$ becomes the leading instability.

If the parameter $\lambda_{1,0}$ takes finite value, the relations between $\lambda_{1}$, $\lambda_{2}$, $\lambda_{3}$, $\lambda_{1}/\lambda_{2}$,
$\lambda_{1}/|\lambda_{3}|$, $\lambda_{2}/|\lambda_{3}|$ and running parameter $\ell$ are shown in Fig.~\ref{Fig:VRGQNLFInilambda1}. $\lambda_{0}$ always equals
to zero, if only $\lambda_{1,0}$ is finite initially. We can find that $\lambda_{1}$
always approaches to infinity at some finite value  $\ell_{c}$. As shown in Fig.~\ref{Fig:VRGQNLFInilambda1}(b), $\lambda_{2}$ flows from zero to infinity
at the same $\ell_{c}$. According to Fig.~\ref{Fig:VRGQNLFInilambda1}(c), $\lambda_{3}$ is generated from zero and approaches to negative infinity finally. As depicted in
Figs.~\ref{Fig:VRGQNLFInilambda1}(d)-\ref{Fig:VRGQNLFInilambda1}(f), $\lambda_{1}/\lambda_{2}$ flows to 1, and $\lambda_{1}/|\lambda_{3}|$ and $\lambda_{2}/|\lambda_{3}|$ flow to a
constant smaller than 1 for $N=1$ but flow to a constant larger than 1 for $N\ge2$.
These results represent that transition into superconducting phase is the leading instability
for $N=1$, but generation of $\Delta_{1}$ or $\Delta_{2}$ and splitting of quadratic nodal
line into conventional nodal lines is the leading instability for $N\ge2$.

\begin{figure}[htbp]
\center
\includegraphics[width=3.1in]{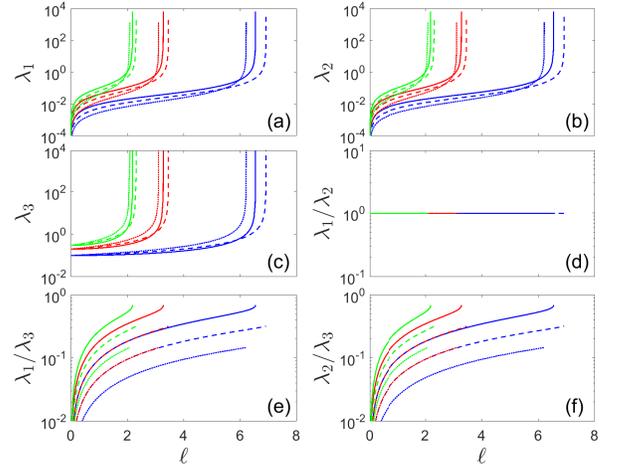}
\caption{Flows of $\lambda_{1}$, $\lambda_{2}$, $\lambda_{3}$, $\lambda_{1}/\lambda_{2}$, $\lambda_{1}/\lambda_{3}$,
and $\lambda_{2}/\lambda_{3}$ in quadratic NLF system. Blue, red, and green colors correspond to $\lambda_{3,0}=0.1, 0.2, 0.3$ respectively.
Solid, dashed, and dotted lines stand for
$N=1, 2, 4$ respectively. \label{Fig:VRGQNLFInilambda3}}
\end{figure}

If the parameter $\lambda_{2,0}$ takes finite value, we will obtain qualitatively similar results comparing the ones in the case that only $\lambda_{1,0}$
is finite.

If the initial value $\lambda_{3,0}$ is finite, the flows of $\lambda_{1}$, $\lambda_{2}$, $\lambda_{3}$, $\lambda_{1}/\lambda_{2}$, $\lambda_{1}/\lambda_{3}$, $\lambda_{2}/\lambda_{3}$ are shown in Fig.~\ref{Fig:VRGQNLFInilambda3}. According to
Figs.~\ref{Fig:VRGQNLFInilambda3}(a)-\ref{Fig:VRGQNLFInilambda3}(c), $\lambda_{3}$ approaches to infinity at a finite $\ell_{c}$, and $\lambda_{1}$ and $\lambda_{2}$ flow
from zero and approach to infinity at the same $\ell_{c}$. As shown in Fig.~\ref{Fig:VRGQNLFInilambda3}(d), $\lambda_{1}/\lambda_{2}$  equals to $1$. As
displayed in Figs.~\ref{Fig:VRGQNLFInilambda3}(e) and \ref{Fig:VRGQNLFInilambda3}(f),
$\lambda_{1}/\lambda_{3}$ and $\lambda_{2}/\lambda_{3}$ always flow to a constant smaller than 1 for any fermion flavor $N$. It represents that generation of excitonic gap is always the
leading instability for any fermion flavor, if only $\lambda_{3,0}$ takes finite value.

\begin{figure}[htbp]
\center
\includegraphics[width=3.1in]{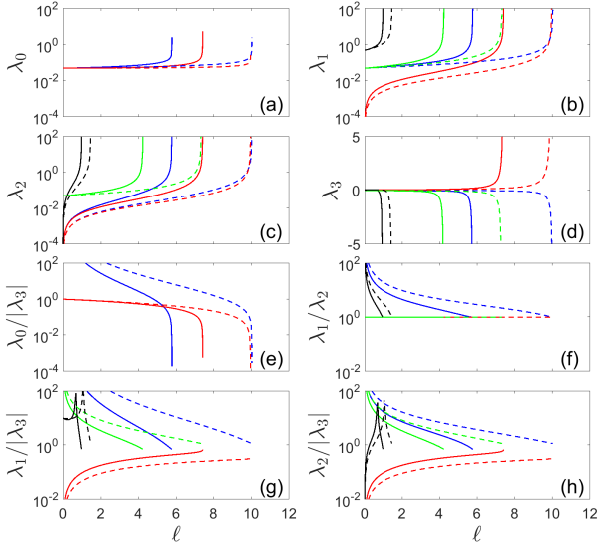}
\caption{Flows of $\lambda_{0}$, $\lambda_{1}$, $\lambda_{2}$, $\lambda_{3}$ $\lambda_{0}/|\lambda_{3}|$, $\lambda_{1}/\lambda_{2}$, $\lambda_{1}/|\lambda_{3}|$,
and $\lambda_{2}/|\lambda_{3}|$ in quadratic NLF system. Blue color corresponds to $\lambda_{0,0}=0.05$ and $\lambda_{1,0}=0.05$;
Red color corresponds to $\lambda_{0,0}=0.05$ and $\lambda_{3,0}=0.05$; Green color corresponds to $\lambda_{1,0}=0.05$ and $\lambda_{2,0}=0.05$;
Black color corresponds to $\lambda_{1,0}=0.05$ and $\lambda_{3,0}=0.05$; Solid and  dashed lines stand for
$N=1$ and $N=2$ respectively. \label{Fig:VRGQNLFMixture}}
\end{figure}

If the initial values of two coupling parameters are finite, the flows of $\lambda_{0}$, $\lambda_{1}$,
$\lambda_{2}$, $\lambda_{3}$, $\lambda_{0}/|\lambda_{3}|$, $\lambda_{1}/\lambda_{2}$, $\lambda_{1}/|\lambda_{3}|$, $\lambda_{2}/|\lambda_{3}|$ are displayed in Fig.~\ref{Fig:VRGQNLFMixture}.
According to these results, we find that the system could be driven to NLSM, excitonic insulator, or superconducting phase, which is determined by
the concrete initial conditions and fermion flavor sensitively.

\subsection{Cubic NLF \label{subsection:RGResultsCubicNLF}}

For cubic NLF system in the presence of four-fermion interactions, the RG equations for the four-fermion coupling parameters are given by
\begin{eqnarray}
\frac{d\lambda_{0}}{d\ell}&=&\frac{1}{3}\lambda_{0}+
\left(\lambda_{1}\lambda_{3}+\lambda_{2}\lambda_{3}\right)\frac{1}{N}, \label{Eq:RGElambda0CNLSM}
\\
\frac{d\lambda_{1}}{d\ell}&=&\frac{1}{3}\lambda_{1}+\bigg[\left(-\lambda_{1}^{2}+\lambda_{0}\lambda_{1}+\lambda_{0}\lambda_{3}
+\lambda_{1}\lambda_{2}+\lambda_{1}\lambda_{3}\right.\nonumber
\\
&&\left.-2\lambda_{2}\lambda_{3}\right)\frac{1}{N}+\lambda_{1}^{2}\bigg] \label{Eq:RGElambda1CNLSM}
,
\\
\frac{d\lambda_{2}}{d\ell}&=&\frac{1}{3}\lambda_{2}+\bigg[\left(-\lambda_{2}^{2}+\lambda_{0}\lambda_{2}+\lambda_{0}\lambda_{3}+\lambda_{1}\lambda_{2}
-2\lambda_{1}\lambda_{3}\right.\nonumber
\\
&&\left.+\lambda_{2}\lambda_{3}\right)\frac{1}{N}+\lambda_{2}^{2}\bigg], \label{Eq:RGElambda2CNLSM}
\\
\frac{d\lambda_{3}}{d\ell}&=&\frac{1}{3}\lambda_{3}+\bigg[\left(-2\lambda_{3}^{2}+\lambda_{0}\lambda_{1}+\lambda_{0}\lambda_{2}+2\lambda_{0}\lambda_{3}
-2\lambda_{1}\lambda_{2}\right.\nonumber
\\
&&\left.+2\lambda_{1}\lambda_{3}+2\lambda_{2}\lambda_{3}\right)\frac{1}{N}
+2\lambda_{3}^{2}\bigg].  \label{Eq:RGElambda3CNLSM}
\end{eqnarray}
The transformations  $\frac{k_{F}}{2\pi B\Lambda}\lambda_{i}\rightarrow\lambda_{i}$ with
$i=0, 1, 2, 3$ have been utilized. One could find that one type of four-fermion interaction can exist solely.

If only consider the four-fermion interaction $\lambda_{0}\left(\psi^{\dag}\sigma_{0}\psi\right)^{2}$,
RG equation for the coupling strength takes the form
\begin{eqnarray}
\frac{d\lambda_{0}}{d\ell}&=&\frac{1}{3}\lambda_{0}.
\end{eqnarray}
The solution is
\begin{eqnarray}
\lambda_{0}=\lambda_{0,0}e^{\frac{1}{3}\ell}.
\end{eqnarray}
It is easy to find that $\lambda_{0}$ does not become divergent at a
finite energy scale, but only becomes divergent in the lowest energy limit $\ell\rightarrow\infty$, i.e, $\ell_{c}\rightarrow\infty$.
We believe that divergence  of four-fermion coupling strength at $\ell\rightarrow\infty$ does not represent the generation of a finite expectation
value of order parameter. Indeed, according to Eqs.~(\ref{Eq:LambdaUS}) and (\ref{Eq:DeltaValue}),
the energy scale for the appearance of instability and the magnitude of order parameter vanish if $\ell_{c}\rightarrow\infty$.

If only the four-fermion interaction $\lambda_{1}\left(\psi^{\dag}\sigma_{1}\psi\right)^{2}$ is considered,
the RG equation for $\lambda_{1}$ is given by
\begin{eqnarray}
\frac{d\lambda_{1}}{d\ell}=\frac{1}{3}\lambda_{1}+\left(1-\frac{1}{N}\right)\lambda_{1}^{2}. \label{Eq:CQLFOnlylambda1}
\end{eqnarray}
For the fermion flavor $N=1$, the RG equation becomes
\begin{eqnarray}
\frac{d\lambda_{1}}{d\ell}=\frac{1}{3}\lambda_{1}.
\end{eqnarray}
The corresponding solution reads as
\begin{eqnarray}
\lambda_{1}&=&\lambda_{1,0}e^{\frac{1}{3}\ell},
\end{eqnarray}
which becomes divergent in the lowest energy limit $\ell\rightarrow\infty$.
In this case, divergence of $\lambda_{1}$ does not indicate the generation of long-range order parameter
$\Delta_{1}=\left<\psi^{\dag}\sigma_{1}\psi\right>$.

\begin{figure}[htbp]
\center
\includegraphics[width=3.1in]{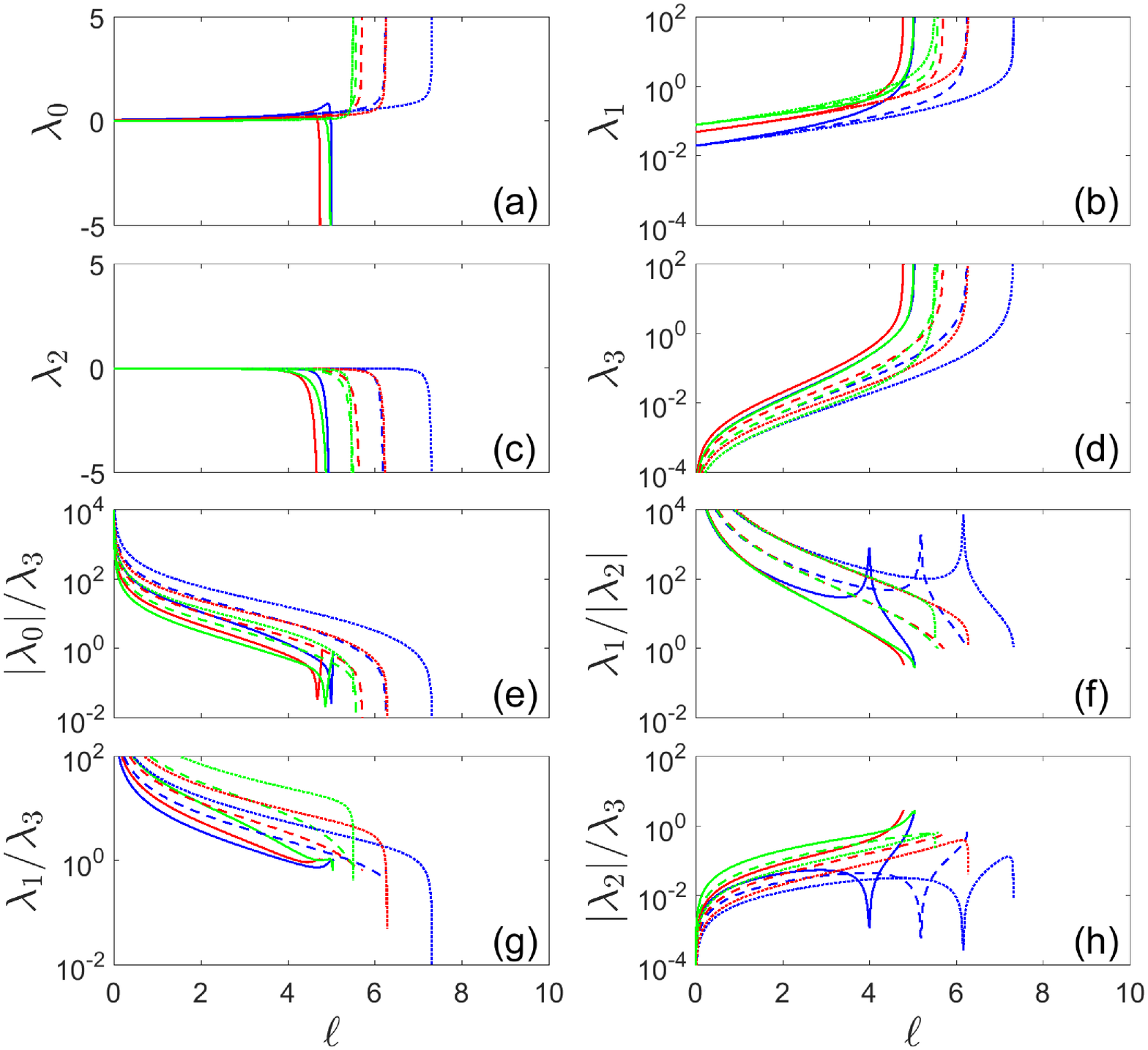}
\caption{Flows of $\lambda_{0}$, $\lambda_{1}$, $\lambda_{2}$, $\lambda_{3}$, $|\lambda_{0}|/\lambda_{3}$, $\lambda_{1}/|\lambda_{2}|$, $\lambda_{1}/\lambda_{3}$
and $|\lambda_{2}|/\lambda_{3}$ in cubic NLF system. Blue color corresponds to $\lambda_{0,0}=0.08$ and $\lambda_{1,0}=0.02$;
Red color corresponds to $\lambda_{0,0}=0.05$ and $\lambda_{1,0}=0.05$;
Green color corresponds to $\lambda_{0,0}=0.02$ and $\lambda_{1,0}=0.08$. Solid, dashed, and dotted lines stand for
$N=1, 2, 4$ respectively. \label{Fig:VRGCNLFMixtureA}}
\end{figure}

For the case $N>1$, solving Eq.~(\ref{Eq:CQLFOnlylambda1}) yields
\begin{eqnarray}
\lambda_{1}&=&\frac{\frac{1}{3}\lambda_{1,0}e^{\frac{1}{3}\ell}}
{\frac{1}{3}+\left(1-\frac{1}{N}\right)\lambda_{1,0}\left(1-e^{\frac{1}{3}\ell}\right)}.
\end{eqnarray}
We find that $\lambda_{1}$ becomes divergent at a critical value
\begin{eqnarray}
\ell_{1c}=3\ln\left[1+\frac{1}{3\left(1-\frac{1}{N}\right)\lambda_{1,0}}\right].
\end{eqnarray}
This result implies that a finite expectation value $\Delta_{1}=\left<\psi^{\dag}\sigma_{1}\psi\right>$ is
generated under the influence of four-fermion interaction. The magnitude of $\Delta_{1}$ can be roughly estimated by
\begin{eqnarray}
\Delta_{1}\sim\Lambda e^{-\ell_{1c}},
\end{eqnarray}
where $\Lambda$ is an energy cutoff.

Similarly, if only the four-fermion interaction $\lambda_{2}\left(\psi^{\dag}\sigma_{2}\psi\right)^{2}$ is
considered, we could
find that a finite expectation value $\Delta_{2}=\left<\psi^{\dag}\sigma_{2}\psi\right>$  is generated for $N>1$.

Considering only the four-fermion interaction $\lambda_{3}\left(\psi^{\dag}\sigma_{3}\psi\right)^{2}$, the RG equation for the
coupling strength can be written as
\begin{eqnarray}
\frac{d\lambda_{3}}{d\ell}&=&\frac{1}{3}\lambda_{3}+2\left(1-\frac{1}{N}\right)\lambda_{3}^{2}.  \label{Eq:CQLFOnlylambda3}
\end{eqnarray}
If $N=1$, the RG equation becomes
\begin{eqnarray}
\frac{d\lambda_{3}}{d\ell}&=&\frac{1}{3}\lambda_{3}.
\end{eqnarray}
The corresponding solution is given by
\begin{eqnarray}
\lambda_{3}=\lambda_{3,0}e^{\frac{1}{3}\ell},
\end{eqnarray}
which is divergent in the limit $\ell\rightarrow\infty$. Thus, for $N=1$, finite expectation value
$\Delta_{3}=\left<\psi^{\dag}\sigma_{3}\psi\right>$ should not be generated.

\begin{figure}[htbp]
\center
\includegraphics[width=3.1in]{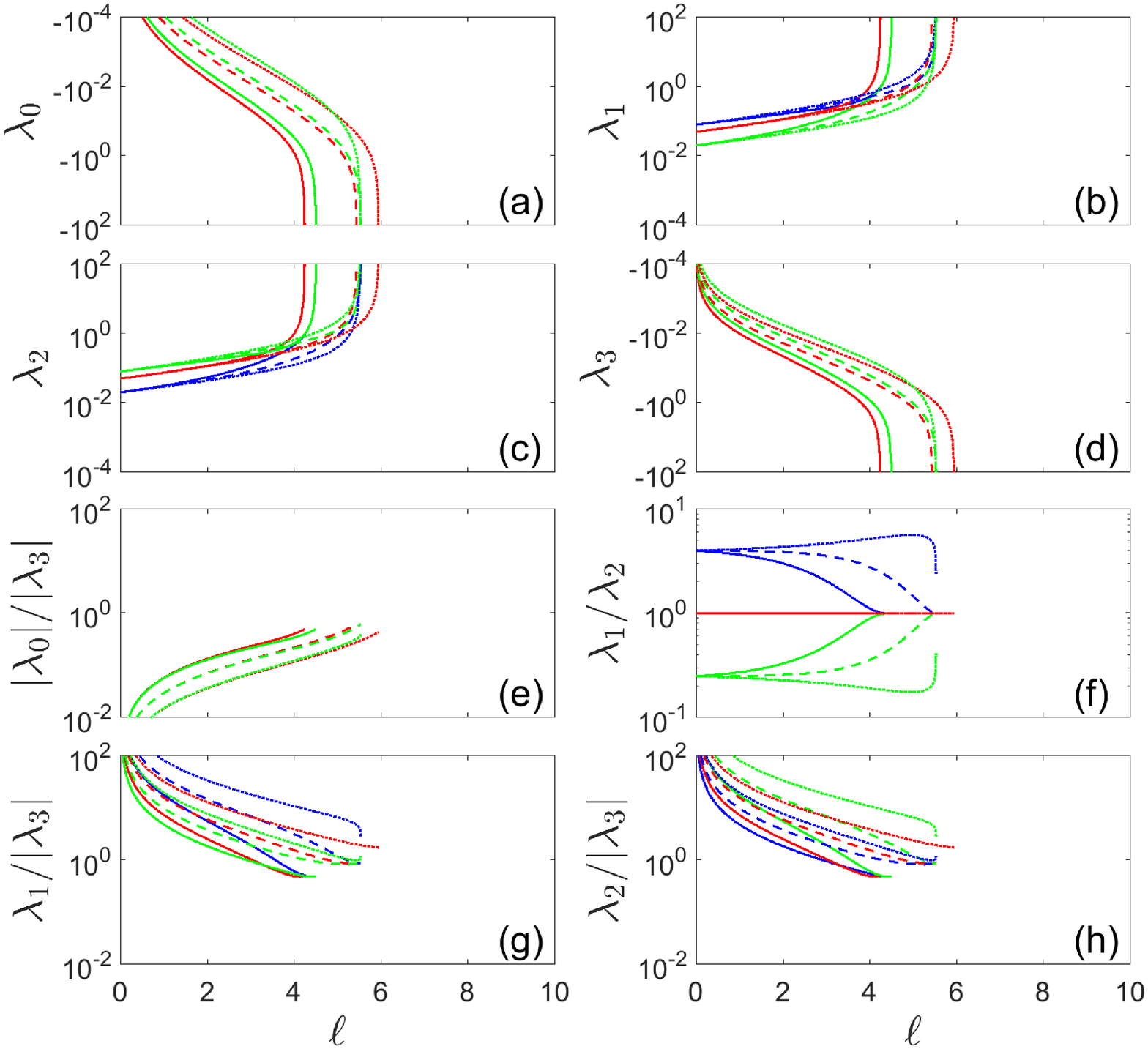}
\caption{Flows of $\lambda_{0}$, $\lambda_{1}$, $\lambda_{2}$, $\lambda_{3}$, $|\lambda_{0}|/|\lambda_{3}|$ $\lambda_{1}/\lambda_{2}$, $\lambda_{1}/|\lambda_{3}|$,
and $\lambda_{2}/|\lambda_{3}|$ in cubic NLF system. Blue color corresponds to $\lambda_{1,0}=0.08$ and $\lambda_{2,0}=0.02$;
Red color corresponds to $\lambda_{1,0}=0.05$ and $\lambda_{2,0}=0.05$;
Green color corresponds to $\lambda_{1,0}=0.02$ and $\lambda_{2,0}=0.08$.
Solid, dashed, and dotted lines stand for $N=1, 2, 4$ respectively.
\label{Fig:VRGQNLFMixtureB}}
\end{figure}

For $N>1$, solving Eq.~(\ref{Eq:CQLFOnlylambda3}) gives
rise to
\begin{eqnarray}
\lambda_{3}&=&\frac{\frac{1}{3}\lambda_{3,0}e^{\frac{1}{3}\ell}}{\frac{1}{3}+2\left(1-\frac{1}{N}\right)\lambda_{3,0}
\left(1-e^{\frac{1}{3}\ell}\right)},
\end{eqnarray}
It is found that $\lambda_{3}$ approaches to infinity when $\ell\rightarrow\ell_{3c}$, where
\begin{eqnarray}
\ell_{3c}=3\ln\left[1+\frac{1}{6\left(1-\frac{1}{N}\right)\lambda_{3,0}}\right].
\end{eqnarray}
Therefore, for $N>1$, finite expectation value $\Delta_{3}$ should be generated under the influence of four-fermion interaction $\lambda_{3}\left(\psi^{\dag}\sigma_{3}\psi\right)^{2}$.
The magnitude of $\Delta_{3}$ can be estimated by
\begin{eqnarray}
\Delta_{3}\sim\Lambda e^{-\ell_{3c}}.
\end{eqnarray}

A shown in above, if one type of four-fermion coupling is considered and $N=1$,
long-range order is  not generated in cubic NLF system.
Accordingly, the DOS takes the behavior $\rho(\omega)\sim\omega^{-1/3}$ which is divergent in the limit $\omega\rightarrow0$,
and compressibility takes the behavior $C_{v}(T)\sim T^{-1/3}$ which is divergent in the limit $T\rightarrow0$, as shown in the Sec.~\ref{Sec:ObserQuant}.
We notice that these characteristics are similar to ones in  supermetal state
proposed by Isobe and Fu \cite{Isobe19}. Indeed, as shown in the paper by Isobe and Fu, in the supermetal state,
none long-range order appears, DOS becomes divergent in the limit $\omega\rightarrow0$, and compressibility
becomes divergent in the limit $T\rightarrow0$.

\begin{figure}[htbp]
\center
\includegraphics[width=3.1in]{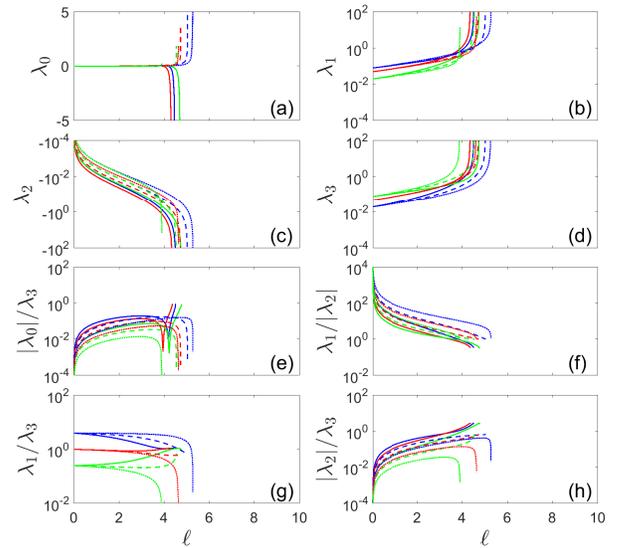}
\caption{Flows of $\lambda_{0}$, $\lambda_{1}$, $\lambda_{2}$, $\lambda_{3}$, $|\lambda_{0}|/\lambda_{3}$, $\lambda_{1}/|\lambda_{2}|$, $\lambda_{1}/\lambda_{3}$,
and $|\lambda_{2}|/\lambda_{3}$ in cubic NLF system. Blue color corresponds to $\lambda_{1,0}=0.08$ and $\lambda_{3,0}=0.02$;
Red color corresponds to $\lambda_{1,0}=0.05$ and $\lambda_{3,0}=0.05$;
Green color corresponds to $\lambda_{1,0}=0.02$ and $\lambda_{3,0}=0.08$. Solid, dashed, and dotted lines stand for
$N=1, 2, 4$ respectively. \label{Fig:VRGCNLFMixtureC}}
\end{figure}

If the initial values of two coupling parameters are finite, the flows of $\lambda_{0}$, $\lambda_{1}$,
$\lambda_{2}$, $\lambda_{3}$, and the ratios between them are shown in Figs.~\ref{Fig:VRGCNLFMixtureA}-\ref{Fig:VRGCNLFMixtureC}.
We notice that the coupling parameters which vanish initially are generated. The absolute values of $\lambda_{0}$, $\lambda_{1}$, $\lambda_{2}$, and
$\lambda_{3}$  all approach to infinity at a finite RG running parameter $\ell_{c}$. The ratios between
the coupling parameters in the limit $\ell\rightarrow\ell_{c}$ are determined by the initial conditions.
After checking these ratios, we find that the system would become to NLSM, excitonic insulator, or superconducting phase,
according to the concrete initial conditions and the value of $N$.

\section{Observable quantities \label{Sec:ObserQuant}}

For convenience, we compare the observable quantities in different phases.

For conventional NLF system, the DOS satisfies
\begin{eqnarray}
\rho(\omega) = \frac{Nk_{F}|\omega|}{2\pi v_{F}v_{z}},
\end{eqnarray}
where $v_{F}$ and $v_{z}$ are the fermion velocities
within the $x$-$y$ plane and along the $z$ axis.
The specific heat and compressibility depend on temperature as
\begin{eqnarray}
C_{v}(T) &=& \frac{9\zeta(3)Nk_{F}}{\pi v_{F}v_{z}}T^{2},
\\
\kappa(T) &=& \frac{2\ln(2)Nk_{F}}{\pi v_{F}v_{z}}T,
\end{eqnarray}
where $\zeta(x)$ is the Riemann zeta function.

For quadratic and cubic NLF systems with an excitonic gap $\Delta_{3}$,  the retarded fermion propagator takes the form
\begin{eqnarray}
G_{q,c}^{\mathrm{ret}}(\omega,\mathbf{k})=\frac{1}{-\left(\omega+i\eta\right)
+\mathcal{H}_{0}^{q,c}(\mathbf{k})+\Delta_{3}\sigma_{3}},
\end{eqnarray}
where $\eta$ is infinitesimal.
The spectral function is given by
\begin{eqnarray}
\mathcal{A}_{q,c}(\omega,\mathbf{k})&=&\frac{1}{\pi}\mathrm{Tr}\left[\mathrm{Im}\left[G_{q,c}^{\mathrm{ret}}(\omega,\mathbf{k})\right]\right]\nonumber
\\
&=&2|\omega|\delta\left(\omega^{2}-\left(E_{q,c}^{2}(\mathbf{k})
+\Delta_{3}^{2}\right)\right).\label{Eq:SpectralFunction}
\end{eqnarray}
The DOS can be written as
\begin{eqnarray}
\rho_{q,c}(\omega)&=&N\int\frac{d^3\mathbf{k}}{(2\pi)^{3}}\mathcal{A}_{q,c}(\omega,\mathbf{k})\nonumber
\\
&\approx&Nk_{F}\int\frac{dk_{r}dk_{z}}{(2\pi)^{2}}\mathcal{A}_{q,c}(\omega,\mathbf{k}). \label{Eq:DOSExpression}
\end{eqnarray}
Substituting Eq.~(\ref{Eq:DispersionQNLF}) into Eqs.~(\ref{Eq:SpectralFunction}) and (\ref{Eq:DOSExpression}), we can get the DOS for
quadratic NLF system
\begin{eqnarray}
\rho_{q}(\omega)&=&\frac{Nk_{F}|\omega|}{4\pi A\sqrt{\omega^{2}-\Delta_{3}^{2}}}\theta\left(\left|\omega\right|-\left|\Delta_{3}\right|\right),
\end{eqnarray}
where $\theta(x)$ represents the Heaviside step function.
Substituting Eq.~(\ref{Eq:DispersionCNLF}) into Eqs.~(\ref{Eq:SpectralFunction}) and (\ref{Eq:DOSExpression}), the DOS for
cubic NLF system can be written as
\begin{eqnarray}
\rho_{c}(\omega)
&=&\frac{Nk_{F}|\omega|}{6\pi B^{2/3}\left(\omega^{2}-\Delta_{3}^{2}\right)^{2/3}}
\theta\left(\left|\omega\right|-\left|\Delta_{3}\right|\right).
\end{eqnarray}
If $\Delta_{3}=0$, $\rho_{q}$ and $\rho_{c}$ become
\begin{eqnarray}
\rho_{q}(\omega)&=&\frac{Nk_{F}}{4\pi A},
\end{eqnarray}
and
\begin{eqnarray}
\rho_{c}(\omega)
&=&\frac{Nk_{F}}{6\pi B^{2/3}|\omega|^{1/3}},
\end{eqnarray}
respectively.

For quadratic and cubic NLF systems with finite excitonic gap $\Delta_{3}$ and finite chemical potential $\mu$,
the propagator of fermions in the Matsubara formalism can be written as
\begin{eqnarray}
G_{q,c}(\omega_{n},\mathbf{k})&=&\frac{1}{-\left(i\omega_{n}+\mu\right)
+\mathcal{H}_{0}^{q,c}(\mathbf{k})+\Delta_{3}\sigma_{3}}\nonumber
\\
&=&\frac{i\omega_{n}+\mu
+\mathcal{H}_{0}^{q,c}(\mathbf{k})+\Delta_{3}\sigma_{3}}{\left(\omega_{n}-i\mu\right)^{2}
+E_{q,c}^{2}(\mathbf{k})+\Delta_{3}^{2}},
\end{eqnarray}
where $\omega_{n}=(2n+1)\pi T$ with $n$ being integers. The parameter chemical potential $\mu$ is introduced to calculate
the compressibility subsequently.
The free energy of the fermions is given by
\begin{eqnarray}
F_{f}(T,\mu)
&=&-2NT\sum_{\omega_{n}}\int\frac{d^3\mathbf{k}}{(2\pi)^3}\nonumber
\\
&&\times\ln\left[\left(\left(\omega_{n}-i\mu\right)^2+E_{q,c}'^{2}(\mathbf{k})\right)^{1/2}\right],
\end{eqnarray}
where
\begin{eqnarray}
E_{q,c}'(\mathbf{k})=\sqrt{E_{q,c}^{2}(\mathbf{k})+\Delta_{3}^{2}}.
\end{eqnarray}
Carrying out the frequency summation, we obtain
\begin{eqnarray}
F_{f}(T,\mu) &=&-2N\sum_{\alpha=\pm1}\int\frac{d^3\mathbf{k}}{(2\pi)^3}\bigg[E_{q,c}'(\mathbf{k})\nonumber
\\
&&+T\ln\left(1+e^{-\frac{E_{q,c}'(\mathbf{k})+\alpha\mu}{T}}\right)\bigg],
\end{eqnarray}
which is clearly divergent. In order to get a finite free energy, we
redefine $F_{f}(T)-F_{f}(0)$ as $F_{f}(T)$, and get
\begin{eqnarray}
F_{f}(T,\mu) &=&-2NT\sum_{\alpha=\pm1}\int\frac{d^3\mathbf{k}}{(2\pi)^3}
\ln\left(1+e^{-\frac{E_{q,c}'(\mathbf{k})\pm\alpha\mu}{T}}\right)\nonumber
\\
&\approx&-2NTk_{F}\sum_{\alpha=\pm1}\int\frac{dk_{r}dk_{z}}{(2\pi)^2}\nonumber
\\
&&\times\ln\left(1+e^{-\frac{E_{q,c}'(\mathbf{k})\pm\alpha\mu}{T}}\right). \label{Eq:FreeEnergyFiniteChemicalPotential}
\end{eqnarray}

Taking the limit $\mu=0$, we have
\begin{eqnarray}
F_{f}(T)
&=&-4NTk_{F}\int\frac{dk_{r}dk_{z}}{(2\pi)^2}\ln\left(1+e^{-\frac{E_{q,c}'(\mathbf{k})}{T}}\right). \label{Eq:FreeEnergyZeroChemicalPotential}
\end{eqnarray}
The specific heat is defined as
\begin{eqnarray}
C_{v}(T)&=&-T\frac{\partial^2 F_{f}(T)}{\partial T^2}. \label{Eq:CvDefinition}
\end{eqnarray}
Substituting Eq.~(\ref{Eq:DispersionQNLF}) into Eqs.~(\ref{Eq:FreeEnergyZeroChemicalPotential}) and (\ref{Eq:CvDefinition}), we find that for quadratic
NLF system, if $\Delta_{3}=0$, the specific heat reads as
\begin{eqnarray}
C_{v}(T)
&=&\frac{\pi Nk_{F}}{24A}T;
\end{eqnarray}
If $\Delta_{3}$ is finite, the specific heat satisfies
\begin{eqnarray}
C_{v}(T)&\approx&\frac{Nk_{F}}{\pi A}\frac{\Delta_{3}^{3}}{T^{2}}
e^{-\frac{\Delta_{3}}{T}},
\end{eqnarray}
in the limit $T\ll\Delta_{3}$.
For cubic NLF system, substituting Eq.~(\ref{Eq:DispersionCNLF}) into Eqs.~(\ref{Eq:FreeEnergyZeroChemicalPotential}) and (\ref{Eq:CvDefinition}),
if $\Delta_{3}=0$, we obtain
\begin{eqnarray}
C_{v}(T)
&=&\frac{20a_{1}NT^{2/3}k_{F}}{9\pi B^{2/3}},
\end{eqnarray}
where
\begin{eqnarray}
a_{1}=\int_{0}^{+\infty} dxx
\ln\left(1+e^{-x^{3}}\right)\approx0.3547;
\end{eqnarray}
For finite $\Delta_{3}$, we have
\begin{eqnarray}
C_{v}(T)&\approx&\frac{Nk_{F}}{\pi B^{2/3}}\frac{\Delta_{3}^{8/3}}{T^{2}}
e^{-\frac{\Delta_{3}}{T}},
\end{eqnarray}
in the limit $T\ll\Delta_{3}$.

The compressibility is defined as
\begin{eqnarray}
\kappa(T,\mu)=-\frac{\partial^2 F_{f}(T,\mu)}{\partial \mu^2}. \label{Eq:CompressiblityDefinition}
\end{eqnarray}
Substituting Eq.~(\ref{Eq:FreeEnergyFiniteChemicalPotential}) into Eq.~(\ref{Eq:CompressiblityDefinition})
and then taking $\mu=0$, we can get the expressions of compressibility for quadratic and cubic NLF systems.
Concretely, for quadratic NLF system, the compressibility is given by
\begin{eqnarray}
\kappa(T)&=&\frac{Nk_{F}}{2\pi A },
\end{eqnarray}
in the case $\Delta_{3}=0$, and
\begin{eqnarray}
\kappa(T)&\approx&\frac{Nk_{F}}{\pi A}\frac{\Delta_{3}^{2}}{T^{2}}
e^{-\frac{\Delta_{3}}{T}},
\end{eqnarray}
for finite $\Delta_{3}$ in the limit $T\ll \Delta_{3}$. For cubic NLF system, in the case $\Delta_{3}=0$, the compressibility reads as
\begin{eqnarray}
\kappa(T)
&=&\frac{2a_{2}Nk_{F}}{\pi B^{2/3} }T^{-1/3}, \label{Eq:KappaCubicNLF}
\end{eqnarray}
where
\begin{eqnarray}
a_{2}=\int_{0}^{+\infty}dxx
\frac{e^{x^{3}}}{\left(1+e^{x^{3}}\right)^{2}}\approx0.1903.
\end{eqnarray}
As shown in Eq.~(\ref{Eq:KappaCubicNLF}), the compressibility $\kappa$ of cubic NLF system  is
divergent in the limit $T\rightarrow0$. This singular behavior of $\kappa$ is closely related to the singular behavior
of DOS $\rho(\omega)\sim\omega^{-1/3}$, which is divergent in the limit $\omega\rightarrow0$. Divergence of DOS at the Fermi level
indicates that the influence of short-range four-interaction would be remarkable. Indeed, as shown in Sec.~\ref{subsection:RGResultsCubicNLF}, in many cases,
arbitrarily weak four-fermion interactions could drive the system to become unstable to a new phase.
Thus, the singular behavior of $\kappa$ is indeed an indication that
the influence of four-fermion interactions is important in cubic NLF
system.
For finite $\Delta_{3}$,
\begin{eqnarray}
\kappa(T)
&\approx&\frac{Nk_{F}}{\pi B^{2/3}}\frac{\Delta_{3}^{2/3}}{T}
e^{-\frac{\Delta_{3}}{T}},
\end{eqnarray}
in the limit $T\ll\Delta_{3}$.

In the RG analysis, the temperature can be introduced through the transformation
$T=T_{0}e^{-\ell}$, where $T_{0}$ is the initial value of temperature and $\ell$ is the RG running parameter.
For example, this transformation was utilized to calculate the corrections of observable quantities induced by  long-range Coulomb interaction in graphene
\cite{Scheehy07, WangLiu14}, 3D DSM/WSM \cite{Goswami11, Hosur12}, multi-WSMs \cite{Lai15, Jian15, WangLiuZhang17B, ZhangShiXin17},
 through incorporating the renormalization of fermion velocities.

We have found that the four-fermion coupling parameters could become divergent at a critical
value $\ell_{c}$. It represents that  if $T<T_{c}$, where $T_{c}=T_{0}e^{-\ell_{c}}$, the system becomes unstable to a new phase, which may
be conventional NLF phase, excitonic insulating phase, or superconducting phase. For the case $\ell<\ell_{c}$, the four-fermion coupling parameters have not flowed
to the strong-coupling regime. Thus, if $T>T_{c}$, the system is stable and still in the original phase.
Accordingly, the observable quantities take the same forms as the free fermion system, due to that  the fermion dispersion is not
renormalized by the short-range four-fermion interactions in this case. If $T<T_{c}$, the behaviors of
observable quantities are modified obviously, since the system becomes to be in a new phase.

\section{Role of the geometry of the nodal line \label{Sec:GeometryNL}}

Thereinbefore, we consider a nodal line with circular shape. It is interesting to verify whether the results are changed by
the geometry of the nodal line. In this section, we consider a straight nodal line along the $y$ axis from $-\Lambda_{y}$ to $\Lambda_{y}$.
The expressions of Hamiltonian density for quadratic and cubic NLSMs are given by
\begin{eqnarray}
\mathcal{H}_{0}=A\left[\left(k_{x}^{2}-k_{z}^{2}\right)\sigma_{1}
+2k_{x}k_{z}\sigma_{2}\right],
\end{eqnarray}
and
\begin{eqnarray}
\mathcal{H}_{0}=B\left[\left(k_{x}^{3}-3k_{x}k_{z}^{2}\right)\sigma_{1}
+\left(k_{z}^{3}-3k_{z}k_{x}^{2}\right)\sigma_{2}\right],
\end{eqnarray}
respectively. After tedious calculation and derivation, we find that the expressions of the RG equations for the
four-fermion coupling parameters are not changed. For quadratic NLF system, the RG equations are still given by Eqs.~(\ref{Eq:RGElambda0QNLSM})-(\ref{Eq:RGElambda3QNLSM}).   The transformations
\begin{eqnarray}
\frac{\Lambda_{y}}{2\pi^{2} A}\lambda_{i}\rightarrow\lambda_{i},
\end{eqnarray}
with $i=0,1,2,3$, have been used in the derivation.
For cubic NLF system, the RG equations are still given by Eqs.~(\ref{Eq:RGElambda0CNLSM})-(\ref{Eq:RGElambda3CNLSM}).
The transformations
\begin{eqnarray}
\frac{\Lambda_{y}}{2\pi^{2} B\Lambda}\lambda_{i}\rightarrow\lambda_{i},
\end{eqnarray}
with $i=0,1,2,3$, have been utilized in the derivation. Thus, the results shown in former sections are still valid for the system with
straight nodal line. We believe that the results also hold on for a system in which the nodal lines take other shapes.

\section{Summary and discussion \label{Sec:Summary}}

In summary, we study the influence of four-fermion interactions on
the quadratic and cubic NLF systems.  Through RG analysis, we find that
arbitrarily weak four-fermion interactions could drive the system to NLSM, excitonic insulator,
or superconducting phase, which is determined by the concrete initial conditions and
value of fermion flavor. The remarkable interaction effects in quadratic and
cubic NLF  systems are closely related to the dispersion of fermion
excitations.

Yu \emph{et al.} predicted that quadratic  NLF system may be realized in
the candidate materials including ZrPtGa, V$_{12}$P$_{7}$, ZrRuAs, and cubic NLF system may
be realized in CaAgBi \cite{Yu19}. We expect our  theoretical predictions may be verified experimentally in these candidate
materials for quadratic and cubic NLF systems in future.

Recently, Volkov and Moroz found that nodal surface fermion system is another
strong correlated system in three dimension, since it would be driven to
excitonic insulating  phase under
arbitrarily weak Coulomb interaction \cite{Volkov18}.

\section*{ACKNOWLEDGEMENTS}

J.R.W. is grateful to Prof. G.-Z. Liu for the valuable discussions.
We acknowledge the support from the National Key R\&D Program of
China under Grants 2017YFA0403600 and 2016YFA0300404,
the National Natural Science Foundation of China under Grants
11504379, 11674327, 11974356, and U1832209, and the Collaborative
Innovation Program of Hefei Science Center CAS under
Grant 2019HSC-CIP002. A portion of this work
was supported by the High Magnetic Field Laboratory of Anhui Province.

\appendix

\section{From lattice model to low-energy effective model \label{App:LatticeModel}}

In this section, we show the lattice models for quadratic and cubic NLF systems, and
derive the corresponding low-energy effective models.

\subsection{Quadratic NLF system}

We consider a lattice Hamiltonian for quadratic NLF system as following
\begin{eqnarray}
\mathcal{H}^{q}&=&-t_{q}\left[\cos\left(k_{r}\right)-\cos\left(\frac{k_{r}}{2}\right)\cos\left(\frac{\sqrt{3}k_{z}}{2}\right)\right]\sigma_{1}\nonumber
\\
&&+\sqrt{3}t_{q}\sin\left(\frac{k_{r}}{2}\right)\sin\left(\frac{\sqrt{3}k_{z}}{2}\right)\sigma_{2}.
\end{eqnarray}
Expanding it around the nodal line determined by $k_{r}=0,k_{z}=0$, we get
\begin{eqnarray}
\mathcal{H}^{q}&\approx&-t_{q}\left[\left(1-\frac{k_{r}^{2}}{2}\right)-\left(1-\frac{k_{r}^{2}}{8}\right)\left(1-\frac{3k_{z}^{2}}{8}\right)\right]\sigma_{1}\nonumber
\\
&&+\sqrt{3}t_{q}\frac{k_{r}}{2}\frac{\sqrt{3}k_{z}}{2}\sigma_{2}\nonumber
\\
&\approx&\frac{3}{8}t_{q}\left[\left(k_{r}^{2}-k_{z}^{2}\right)\sigma_{1}
+2k_{r}k_{z}\sigma_{2}\right]\nonumber
\\
&=&A\left[\left(k_{r}^{2}-k_{z}^{2}\right)\sigma_{1}
+2k_{r}k_{z}\sigma_{2}\right],
\end{eqnarray}
where
\begin{eqnarray}
A=\frac{3}{8}t_{q}.
\end{eqnarray}

\subsection{Cubic NLF system}

\begin{figure}[htbp]
\center
\includegraphics[width=3.1in]{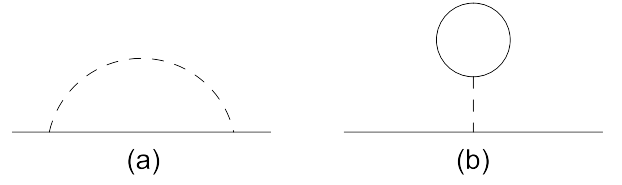}
\caption{Feynman diagrams for the self-energies  of fermions induced by four-fermion interactions. Solid
line represents the fermion propagator, and dashed line stands for the four-fermion interaction.\label{Fig:FermionSelfEnergy}}
\end{figure}

We consider a lattice Hamiltonian for cubic NLF system as following
\begin{eqnarray}
\mathcal{H}^{c}&=&t_{c}\sin\left(\frac{\sqrt{3}k_{r}}{2}\right)\left[\cos\left(\frac{3k_{z}}{2}\right)-\cos\left(\frac{\sqrt{3}k_{r}}{2}\right)\right]\sigma_{1}\nonumber
\\
&&-3\sqrt{3}t_{c}\sin\left(\frac{k_{z}}{2}\right)\left[\cos\left(\frac{k_{z}}{2}\right)-\cos\left(\frac{\sqrt{3}k_{r}}{2}\right)\right]\sigma_{2}.\nonumber
\\
\end{eqnarray}
Expanding it around the nodal line decided by $k_{r}=0,k_{z}=0$, we obtain
\begin{eqnarray}
\mathcal{H}^{c}&\approx&t_{c}\frac{\sqrt{3}k_{r}}{2}\left[\left(1-\frac{9k_{z}^{2}}{8}\right)-\left(1-\frac{3k_{r}^{2}}{8}\right)\right]\sigma_{1}\nonumber
\\
&&-3\sqrt{3}t_{c}\frac{k_{z}}{2}\left[\left(1-\frac{k_{z}^{2}}{8}\right)-\left(1-\frac{3k_{r}^{2}}{8}\right)\right]\sigma_{2}\nonumber
\\
&=&\frac{3\sqrt{3}t_{c}}{16}\left[\left(k_{r}^{3}-3k_{r}k_{z}^{2}\right)\sigma_{1}
+\left(k_{z}^{3}-3k_{z}k_{r}^{2}\right)\sigma_{2}\right]\nonumber
\\
&=&B\left[\left(k_{r}^{3}-3k_{r}k_{z}^{2}\right)\sigma_{1}
+\left(k_{z}^{3}-3k_{z}k_{r}^{2}\right)\sigma_{2}\right],
\end{eqnarray}
where
\begin{eqnarray}
B=\frac{3\sqrt{3}t_{c}}{16}.
\end{eqnarray}

\section{Fermion propagator}

The fermion propagator for quadratic NLF system takes the form
\begin{eqnarray}
G_{q0}(\omega,\mathbf{k})=\frac{1}{-i\omega+\mathcal{H}_{0}^{q}(\mathbf{k})}, \label{Eq:PropagatorQNLF}
\end{eqnarray}
where $\mathcal{H}_{0}^{q}$ is given by Eq.~(\ref{Eq:HamiltonianDensityQNLF}).
The fermion propagator for cubic NLF system can be written as
\begin{eqnarray}
G_{c0}(\omega,\mathbf{k})=\frac{1}{-i\omega
+\mathcal{H}_{0}^{c}(\mathbf{k})},
\label{Eq:PropagatorCNLF}
\end{eqnarray}
where $\mathcal{H}_{0}^{c}$ is expressed by Eq.~(\ref{Eq:HamiltonianDensityCNLF}).

\section{Self-energy of the fermions}

\begin{figure}[htbp]
\center
\includegraphics[width=3.1in]{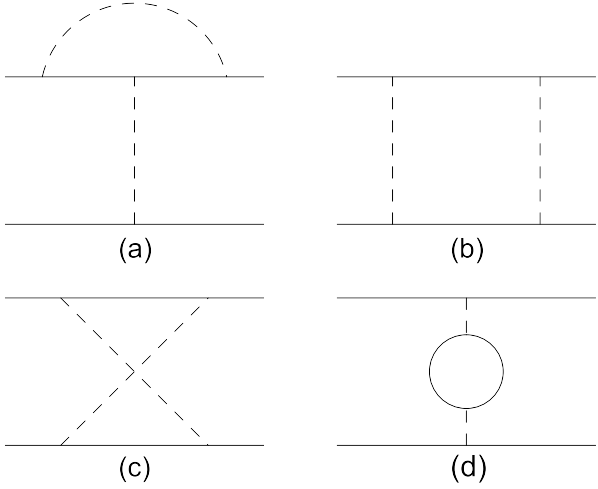}
\caption{One-loop Feynman diagrams for the corrections to the
four-fermion couplings. \label{Fig:VertexCorrection}}
\end{figure}

The self-energy of fermions induced by Fig.~\ref{Fig:FermionSelfEnergy}(a) is defined as
\begin{eqnarray}
\Sigma_{a}=\sum_{i=0}^{3}\frac{\lambda_{i}}{N}\int\frac{d\omega}{2\pi}
\int'\frac{d^3\mathbf{k}}{(2\pi)^{3}}\sigma_{i}G_{q,c0}(\omega,\mathbf{k})\sigma_{i}, \label{Eq:SelfEnergyA}
\end{eqnarray}
where $\int'$ represents that a momentum shell will be properly taken.
Figure~\ref{Fig:FermionSelfEnergy}(b) results in the self-energy of fermions as following
\begin{eqnarray}
\Sigma_{b}=N\sum_{i=0}^{3}\frac{\lambda_{i}}{N}\int\frac{d\omega}{2\pi}
\int'\frac{d^3\mathbf{k}}{(2\pi)^{3}}\mathrm{Tr}\left[G_{q,c0}(\omega,\mathbf{k})\sigma_{i}\right]. \label{Eq:SelfEnergyB}
\end{eqnarray}
Substituting Eq.~(\ref{Eq:PropagatorQNLF}) or Eq.~(\ref{Eq:PropagatorCNLF}) into Eqs.~(\ref{Eq:SelfEnergyA}) and
(\ref{Eq:SelfEnergyB}), we obtain
\begin{eqnarray}
\Sigma_{a}&=&0,
\\
\Sigma_{b}&=&0,
\end{eqnarray}
for both of quadratic and cubic NLF systems. Thus the fermion propagator is not renormalized by the
four-fermion interactions to one-loop order.

\begin{widetext}

\section{One-loop order corrections for the four-fermion couplings \label{App:OneLoopCorrection}}

\subsection{General expressions for the one-loop order corrections }

The correction contributed  by Fig.~\ref{Fig:VertexCorrection}(a) is given by
\begin{eqnarray}
W^{(1)}=\sum_{i=0}^{3}W_{i}^{(1)}, \label{Eq:CorrectionAAll}
\end{eqnarray}
where
\begin{eqnarray}
W_{i}^{(1)}
&=&\sum_{j=0}^{3}\frac{4\lambda_{i}\lambda_{j}}{N}
\left(\psi^{\dag}\sigma_{i}\psi\right)\int\frac{d\omega}{2\pi}\int'\frac{d^3\mathbf{k}}{(2\pi)^{3}}\left[\psi^{\dag}\sigma_{j}G_{q,c0}\left(\omega,\mathbf{k}\right)\sigma_{i}
G_{q,c0}\left(\omega,\mathbf{k}\right)\sigma_{j}\psi\right]. \label{Eq:CorrectionA}
\end{eqnarray}
The diagrams as shown in Figs.~\ref{Fig:VertexCorrection}(b) and \ref{Fig:VertexCorrection}(c) lead to
the correction for the four-fermion couplings as following
\begin{eqnarray}
W^{(2)+(3)}=\sum_{i=0}^{3}\sum_{i\le j\le 3}W_{ij}^{(2)+(3)}, \label{Eq:CorrectionBCAll}
\end{eqnarray}
where
\begin{eqnarray}
W_{ij}^{(2)+(3)}&=&\frac{4\lambda_{i}\lambda_{j}}{N}\int\frac{d\omega}{2\pi}\int'\frac{d^3\mathbf{k}}{(2\pi)^{3}}
\left(\psi^{\dag}\sigma_{i}
G_{q,c0}(\omega,\mathbf{k})\sigma_{j}\psi\right)\left\{\psi^{\dag}
\left[\sigma_{j}G_{q,c0}(\omega,\mathbf{k})\sigma_{i}
+\sigma_{i}G_{q,c0}(-\omega,-\mathbf{k})\sigma_{j}\right]\psi\right\}.
\label{Eq:CorrectionBC}
\end{eqnarray}
The correction for the four-fermion couplings resulting from Fig.~\ref{Fig:VertexCorrection}(d) can be written as
\begin{eqnarray}
W^{(4)}=\sum_{i=0}^{3}W_{i}^{(4)}, \label{Eq:CorrectionDAll}
\end{eqnarray}
where
\begin{eqnarray}
W_{i}^{(4)}
&=&-2\lambda_{i}^{2}\left(\psi^{\dag}\sigma_{i}\psi\right)
\left(\psi^{\dag}\sigma_{i}\psi\right)
\int\frac{d\omega}{2\pi}\int'\frac{d^{3}\mathbf{k}}{(2\pi)^{3}}\mathrm{Tr}\left[\sigma_{i}
G_{q,c0}\left(\omega,\mathbf{k}\right)\sigma_{i}
G_{q,c0}\left(\omega,\mathbf{k}\right)\right]. \label{Eq:CorrectionD}
\end{eqnarray}
A momentum shell $b\Lambda<\sqrt{k_{r}^{2}+k_{z}^{2}}<\Lambda$ with $b=e^{-\ell}$ will be  utilized in
the derivation, where $\ell$ stands for the RG running parameter.

\subsection{Results for quadratic NLF}

Substituting Eq.~(\ref{Eq:PropagatorQNLF}) into Eqs.~(\ref{Eq:CorrectionAAll})-
(\ref{Eq:CorrectionD}), we obtain
\begin{eqnarray}
W^{(1)}
&=&\lambda_{1}\left(\lambda_{0}-\lambda_{1}
+\lambda_{2}+\lambda_{3}\right)\frac{1}{N}\frac{k_{F}}{2\pi A}\ell
\left(\psi^{\dag}\sigma_{1}\psi\right)^{2}\nonumber
+\lambda_{2}\left(\lambda_{0}+\lambda_{1}
-\lambda_{2}+\lambda_{3}\right)\frac{1}{N}\frac{k_{F}}{2\pi A}\ell
\left(\psi^{\dag}\sigma_{2}\psi\right)^{2}\nonumber
\\
&&+\lambda_{3}\left(\lambda_{0}+\lambda_{1}
+\lambda_{2}-\lambda_{3}\right)\frac{1}{N}\frac{k_{F}}{\pi A}\ell
\left(\psi^{\dag}\sigma_{3}\psi\right)^{2}, \label{Eq:W1QNL}
\\
W^{(2)+(3)}
&=&\left(\lambda_{0}\lambda_{1}+\lambda_{0}\lambda_{2}\right)\frac{1}{N}\frac{k_{F}}{2\pi A}\ell
\left(\psi^{\dag}\sigma_{0}
\psi\right)^{2}+\left(\sum_{i=0}^{3}\lambda_{i}^{2}+\lambda_{1}\lambda_{2}-2\lambda_{2}\lambda_{3}\right)\frac{1}{N}
\frac{k_{F}}{2\pi A}\ell\left(\psi^{\dag}
\sigma_{1}\psi\right)^{2}\nonumber
\\
&&+\left(\sum_{i=0}^{3}\lambda_{i}^{2}+\lambda_{1}\lambda_{2}-2\lambda_{1}\lambda_{3}\right)
\frac{1}{N}\frac{k_{F}}{2\pi A}
\ell\left(\psi^{\dag}
\sigma_{2}\psi\right)^{2}+\left(-2\lambda_{1}\lambda_{2}+\lambda_{1}\lambda_{3}+\lambda_{2}\lambda_{3}\right)
\frac{1}{N}\frac{k_{F}}{2\pi A}\ell
\left(\psi^{\dag}\sigma_{3}\psi\right)^{2},  \label{Eq:W23QNL}
\\
W^{(4)}
&=&\lambda_{1}^{2}\frac{k_{F}}{2\pi A}\ell\left(\psi^{\dag}\sigma_{1}\psi\right)^{2}
+\lambda_{2}^{2}\frac{k_{F}}{2\pi A}\ell
\left(\psi^{\dag}\sigma_{2}\psi\right)^{2}
+\lambda_{3}^{2}\frac{k_{F}}{\pi A}\ell
\left(\psi^{\dag}\sigma_{3}\psi\right)^{2}.  \label{Eq:W4QNL}
\end{eqnarray}
From Eqs.~(\ref{Eq:W1QNL})-(\ref{Eq:W4QNL}), we get
\begin{eqnarray}
W=W^{(1)}+W^{(2)+(3)}+W^{(4)}=\sum_{i=0}^{3}\delta\lambda_{i}\left(\psi^{\dag}\sigma_{i}\psi\right)^{2},
\end{eqnarray}
where
\begin{eqnarray}
\delta\lambda_{0}&=&
\left(\lambda_{0}\lambda_{1}+\lambda_{0}\lambda_{2}\right)\frac{1}{N}\frac{k_{F}}{2\pi A}\ell, \label{Eq:deltalambda0QNLF}
\\
\delta\lambda_{1}&=& \left[\left(\lambda_{0}^{2}+\lambda_{2}^{2}
+\lambda_{3}^{2}+
\lambda_{0}\lambda_{1}+2\lambda_{1}\lambda_{2}+\lambda_{1}\lambda_{3}
-2\lambda_{2}\lambda_{3}
\right)\frac{1}{N}+\lambda_{1}^{2}\right]
\frac{k_{F}}{2\pi A}\ell, \label{Eq:deltalambda1QNLF}
\\
\delta\lambda_{2}&=&\left[\left(\lambda_{0}^{2}+\lambda_{1}^{2}
+\lambda_{3}^{2}+\lambda_{0}\lambda_{2}+2\lambda_{1}\lambda_{2}-2\lambda_{1}\lambda_{3}
+\lambda_{2}\lambda_{3}\right)
\frac{1}{N}+\lambda_{2}^{2}\right]\frac{k_{F}}{2\pi A}
\ell, \label{Eq:deltalambda2QNLF}
\\
\delta\lambda_{3}&=&\left[\left(-2\lambda_{3}^{2}+2\lambda_{0}\lambda_{3}-2\lambda_{1}\lambda_{2}+3\lambda_{1}\lambda_{3}+3\lambda_{2}\lambda_{3}
\right)
\frac{1}{N}+2\lambda_{3}^{2}\right]\frac{k_{F}}{2\pi A}\ell. \label{Eq:delta3ambdaQNLF}
\end{eqnarray}

\subsection{Results for cubic NLF}

Substituting Eq.~(\ref{Eq:PropagatorCNLF}) into Eqs.~(\ref{Eq:CorrectionAAll})-
(\ref{Eq:CorrectionD}), we get
\begin{eqnarray}
W^{(1)}
&=&\lambda_{1}\left(\lambda_{0}-\lambda_{1}
+\lambda_{2}+\lambda_{3}\right)\frac{1}{N}\frac{k_{F}}{2\pi B\Lambda}\ell
\left(\psi^{\dag}\sigma_{1}\psi\right)^{2}
+\lambda_{2}\left(\lambda_{0}+\lambda_{1}
-\lambda_{2}+\lambda_{3}\right)\frac{1}{N}\frac{k_{F}}{2\pi B\Lambda}\ell
\left(\psi^{\dag}\sigma_{2}\psi\right)^{2}\nonumber
\\
&&+\lambda_{3}\left(\lambda_{0}+\lambda_{1}
+\lambda_{2}-\lambda_{3}\right)\frac{1}{N}\frac{k_{F}}{\pi B\Lambda}\ell
\left(\psi^{\dag}\sigma_{3}\psi\right)^{2}, \label{Eq:W1CNL}
\\
W^{(2)+(3)}
&=&\left(\lambda_{1}\lambda_{3}+\lambda_{2}\lambda_{3}\right)\frac{1}{N}\frac{k_{F}}{2\pi B\Lambda}\ell
\left(\psi^{\dag}\sigma_{0}\psi\right)^{2}\nonumber
+\left(\lambda_{0}\lambda_{3}-2\lambda_{2}\lambda_{3}\right)\frac{1}{N}\frac{k_{F}}{2\pi B\Lambda}\ell
\left(\psi^{\dag}\sigma_{1}\psi\right)^{2}\nonumber
\nonumber
\\
&&+\left(\lambda_{0}\lambda_{3}-2\lambda_{1}\lambda_{3}\right)\frac{1}{N}\frac{k_{F}}{2\pi B\Lambda}\ell
\left(\psi^{\dag}\sigma_{2}\psi\right)^{2}+\left(\lambda_{0}\lambda_{1}+\lambda_{0}\lambda_{2}-2\lambda_{1}\lambda_{2}\right)\frac{1}{N}
\frac{k_{F}}{2\pi B\Lambda}\ell\left(\psi^{\dag}\sigma_{3}\psi\right)^{2}, \label{Eq:W23CNL}
\\
W^{(4)}
&=&\lambda_{1}^{2}\frac{k_{F}}{2\pi B\Lambda}\ell\left(\psi^{\dag}\sigma_{1}\psi\right)^{2}
+\lambda_{2}^{2}\frac{k_{F}}{2\pi B\Lambda}\ell
\left(\psi^{\dag}\sigma_{2}\psi\right)^{2}
+\lambda_{3}^{2}\frac{k_{F}}{\pi B\Lambda}\ell
\left(\psi^{\dag}\sigma_{3}\psi\right)^{2}. \label{Eq:W4CNL}
\end{eqnarray}
From Eqs.~(\ref{Eq:W1CNL})-(\ref{Eq:W4CNL}), we find
\begin{eqnarray}
W=W^{(1)}+W^{(2)+(3)}+W^{(4)}=\sum_{i=0}^{3}\delta\lambda_{i}\left(\psi^{\dag}\sigma_{i}\psi\right)^{2},
\end{eqnarray}
where
\begin{eqnarray}
\delta\lambda_{0}&=&
\left(\lambda_{1}\lambda_{3}+\lambda_{2}\lambda_{3}\right)\frac{1}{N}\frac{k_{F}}{2\pi B\Lambda}\ell \label{Eq:deltalambda0CNLF},
\\
\delta\lambda_{1}&=& \left[\left(-\lambda_{1}^{2}+\lambda_{0}\lambda_{1}+\lambda_{0}\lambda_{3}
+\lambda_{1}\lambda_{2}+\lambda_{1}\lambda_{3}-2\lambda_{2}\lambda_{3}\right)\frac{1}{N}+\lambda_{1}^{2}\right]
\frac{k_{F}}{2\pi B\Lambda}\ell, \label{Eq:deltalambda1CNLF}
\\
\delta\lambda_{2}&=&\left[\left(-\lambda_{2}^{2}+\lambda_{0}\lambda_{2}+\lambda_{0}\lambda_{3}+\lambda_{1}\lambda_{2}
-2\lambda_{1}\lambda_{3}+\lambda_{2}\lambda_{3}\right)\frac{1}{N}+\lambda_{2}^{2}\right]\frac{k_{F}}{2\pi B\Lambda}\ell, \label{Eq:deltalambda2CNLF}
\\
\delta\lambda_{3}&=&\left[\left(-2\lambda_{3}^{2}+\lambda_{0}\lambda_{1}+\lambda_{0}\lambda_{2}+2\lambda_{0}\lambda_{3}
-2\lambda_{1}\lambda_{2}+2\lambda_{1}\lambda_{3}+2\lambda_{2}\lambda_{3}\right)\frac{1}{N}
+2\lambda_{3}^{2}\right]\frac{k_{F}}{2\pi B\Lambda}\ell.
\label{Eq:deltalambda3CNLF}
\end{eqnarray}

\section{Derivation of the RG equations}

\subsection{Quadratic NLF}

The action for quadratic NLFs is
\begin{eqnarray}
S_{\psi}=\int\frac{d\omega}{2\pi}\frac{d^{3}\mathbf{k}}
{(2\pi)^{3}}\psi^{\dag}(\omega,\mathbf{k})
\left[-i\omega+A\left(k_{r}^{2}-k_{z}^{2}\right)\sigma_{1}
+2Ak_{r}k_{z}\sigma_{2}
\right]\psi(\omega,\mathbf{k}),
\end{eqnarray}
where $k_{r}\equiv k_{\bot}-k_{F}\equiv\sqrt{k_{x}^{2}+k_{y}^{2}}-k_{F}$.
In the low-energy regime, the action for  quadratic NLF can be also written as
\begin{eqnarray}
S_{\psi}\approx k_{F}\int\frac{d\omega}{2\pi}\frac{dk_{r}}{2\pi}\frac{dk_{z}}{2\pi}\psi^{\dag}(\omega,\mathbf{k})
\left[-i\omega+A\left(k_{r}^{2}-k_{z}^{2}\right)\sigma_{1}
+2Ak_{r}k_{z}\sigma_{2}
\right]\psi(\omega,\mathbf{k}).
\end{eqnarray}
Employing the transformations
\begin{eqnarray}
k_{r}&=&k_{r}'e^{-\frac{\ell}{2}}, \label{Eq:krScalingQNLF}
\\
k_{z}&=&k_{z}'e^{-\frac{\ell}{2}}, \label{Eq:kZScalingQNLF}
\\
\omega&=&\omega'e^{-\ell}, \label{Eq:omegaScalingQNLF}
\\
\psi&=&\psi' e^{\frac{3}{2}\ell},  \label{Eq:psiScalingQNLF}
\\
A&=&A', \label{Eq:AScalingQNLF}
\end{eqnarray}
the action of the quadratic NLFs becomes
\begin{eqnarray}
S_{\psi'}=k_{F}\int\frac{d\omega'}{2\pi}\frac{dk_{r}'}{2\pi}
\frac{dk_{z}'}{2\pi}\psi'^{\dag}(\omega',\mathbf{k}')
\left[-i\omega'+A'\left(k_{r}'^{2}-k_{z}'^{2}\right)\sigma_{1}
+2A'k_{r}'k_{z}'\sigma_{2}
\right]\psi'(\omega',\mathbf{k}'),
\end{eqnarray}
which recovers the form of the original action.

The action for the four-fermion interactions between quadratic NLFs is given by
\begin{eqnarray}
S_{\psi^{4}}&=&\frac{1}{N}\sum_{i=0}^{3}\lambda_{i}
\int\frac{d\omega_{1}}{2\pi}\frac{d^{3}\mathbf{k}_{1}}{(2\pi)^{3}}
\frac{d\omega_{2}}{2\pi}\frac{d^{3}\mathbf{k}_{2}}{(2\pi)^{3}}
\frac{d\omega_{3}}{2\pi}\frac{d^{3}\mathbf{k}_{3}}{(2\pi)^{3}}
\psi^{\dag}(\omega_{1},\mathbf{k}_{1})\sigma_{i}
\psi(\omega_{2},\mathbf{k}_{2})\psi^{\dag}(\omega_{3},\mathbf{k}_{3})\sigma_{i}
\psi(\omega_{1}-\omega_{2}+\omega_{3},
\mathbf{k}_{1}-\mathbf{k}_{2}+\mathbf{k}_{3})\nonumber
\\
&\approx&\frac{1}{N}\sum_{i=0}^{3}\lambda_{i}k_{F}^{3}\int
\frac{d\omega_{1}}{2\pi}\frac{dk_{1r}}{2\pi}\frac{dk_{1z}}{2\pi}
\frac{d\omega_{2}}{2\pi}\frac{dk_{2r}}{2\pi}\frac{dk_{2z}}{2\pi}
\frac{d\omega_{3}}{2\pi}\frac{dk_{3r}}{2\pi}\frac{dk_{3z}}{2\pi}
\psi^{\dag}(\omega_{1},\mathbf{k}_{1})\sigma_{i}
\psi(\omega_{2},\mathbf{k}_{2})\psi^{\dag}(\omega_{3},\mathbf{k}_{3})\sigma_{i}\nonumber
\\
&&\times\psi(\omega_{1}-\omega_{2}+\omega_{3},
\mathbf{k}_{1}-\mathbf{k}_{2}+\mathbf{k}_{3}).
\end{eqnarray}
Incorporating the one-loop order corrections, the action becomes
\begin{eqnarray}
S_{\psi^{4}}
&=&\frac{1}{N}\sum_{i=0}^{3}\left(\lambda_{i}+\delta\lambda_{i}\right)k_{F}^{3}\int
\frac{d\omega_{1}}{2\pi}\frac{dk_{1r}}{2\pi}\frac{dk_{1z}}{2\pi}
\frac{d\omega_{2}}{2\pi}\frac{dk_{2r}}{2\pi}\frac{dk_{2z}}{2\pi}
\frac{d\omega_{3}}{2\pi}\frac{dk_{3r}}{2\pi}\frac{dk_{3z}}{2\pi}
\psi^{\dag}(\omega_{1},\mathbf{k}_{1})\sigma_{i}
\psi(\omega_{2},\mathbf{k}_{2})\psi^{\dag}(\omega_{3},\mathbf{k}_{3})\sigma_{i}\nonumber
\\
&&\times\psi(\omega_{1}-\omega_{2}+\omega_{3},
\mathbf{k}_{1}-\mathbf{k}_{2}+\mathbf{k}_{3}).
\end{eqnarray}
Adopting the transformations shown in Eqs.~(\ref{Eq:krScalingQNLF})-(\ref{Eq:psiScalingQNLF}), and
\begin{eqnarray}
\lambda_{i}'&=&\lambda_{i}+\delta\lambda_{i}, \label{Eq:lambdaScalingQNLF}
\end{eqnarray}
the action can be written as
\begin{eqnarray}
S_{\psi'^{4}}
&=&\frac{1}{N}\sum_{i=0}^{3}\lambda_{i}'\int
\frac{d\omega_{1}'}{2\pi}\frac{dk_{1r}'}{2\pi}\frac{dk_{1z}'}{2\pi}
\frac{d\omega_{2}'}{2\pi}\frac{dk_{2r}'}{2\pi}\frac{dk_{2z}'}{2\pi}
\frac{d\omega_{3}'}{2\pi}\frac{dk_{3r}'}{2\pi}\frac{dk_{3z}'}{2\pi}
\psi'^{\dag}(\omega_{1}',\mathbf{k}_{1}')\sigma_{i}
\psi'(\omega_{2}',\mathbf{k}_{2}')\psi'^{\dag}(\omega_{3}',\mathbf{k}_{3}')\sigma_{i}\nonumber
\\
&&\times\psi'(\omega_{1}'-\omega_{2}'+\omega_{3}',
\mathbf{k}_{1}'-\mathbf{k}_{2}'+\mathbf{k}_{3}'),
\end{eqnarray}
which recovers the original form of the action. From Eq.~(\ref{Eq:lambdaScalingQNLF}),
we obtain
\begin{eqnarray}
\frac{d\lambda_{i}}{d\ell}&=&\frac{d\delta\lambda_{i}}{d\ell}. \label{Eq:RGElambdaGeneralQNLF}
\end{eqnarray}
Substituting Eqs.~(\ref{Eq:deltalambda0QNLF})-(\ref{Eq:delta3ambdaQNLF}) into
Eq.~(\ref{Eq:RGElambdaGeneralQNLF}), we get the RG equations
\begin{eqnarray}
\frac{d\lambda_{0}}{d\ell}&=&
\left(\lambda_{0}\lambda_{1}+\lambda_{0}\lambda_{2}\right)\frac{1}{N},
\\
\frac{d\lambda_{1}}{d\ell}&=& \left(\lambda_{0}^{2}+\lambda_{2}^{2}
+\lambda_{3}^{2}+
\lambda_{0}\lambda_{1}+2\lambda_{1}\lambda_{2}+\lambda_{1}\lambda_{3}
-2\lambda_{2}\lambda_{3}
\right)\frac{1}{N}+\lambda_{1}^{2},
\\
\frac{d\lambda_{2}}{d\ell}&=&\left(\lambda_{0}^{2}+\lambda_{1}^{2}
+\lambda_{3}^{2}+\lambda_{0}\lambda_{2}+2\lambda_{1}\lambda_{2}-2\lambda_{1}\lambda_{3}
+\lambda_{2}\lambda_{3}\right)
\frac{1}{N}+\lambda_{2}^{2},
\\
\frac{d\lambda_{3}}{d\ell}&=&\left(-2\lambda_{3}^{2}+2\lambda_{0}\lambda_{3}-2\lambda_{1}\lambda_{2}+3\lambda_{1}\lambda_{3}+3\lambda_{2}\lambda_{3}
\right)
\frac{1}{N}+2\lambda_{3}^{2}.
\end{eqnarray}
The transformations
\begin{eqnarray}
\frac{k_{F}}{2\pi A}\lambda_{i}\rightarrow\lambda_{i},
\end{eqnarray}
with $i=0, 1, 2, 3$ have been used.

\subsection{Cubic NLF}

The action for cubic NLFs takes the form
\begin{eqnarray}
S_{\psi}=\int\frac{d\omega}{2\pi}\frac{d^{3}\mathbf{k}}
{(2\pi)^{3}}\psi^{\dag}(\omega,\mathbf{k})
\left[-i\omega+B\left(k_{r}^{3}-3k_{r}k_{z}^{2}\right)\sigma_{1}
+B\left(k_{z}^{3}-3k_{r}k_{z}^{2}\right)\sigma_{2}
\right]\psi(\omega,\mathbf{k}),
\end{eqnarray}
which is equivalent to
\begin{eqnarray}
S_{\psi}\approx k_{F}\int\frac{d\omega}{2\pi}\frac{dk_{r}}{2\pi}\frac{dk_{z}}{2\pi}\psi^{\dag}(\omega,\mathbf{k})
\left[-i\omega+B\left(k_{r}^{3}-3k_{r}k_{z}^{2}\right)\sigma_{1}
+B\left(k_{z}^{3}-3k_{r}k_{z}^{2}\right)\sigma_{2}
\right]\psi(\omega,\mathbf{k}).
\end{eqnarray}
Using the transformations
\begin{eqnarray}
k_{r}&=&k_{r}'e^{-\frac{\ell}{3}}, \label{Eq:krScalingCNLF}
\\
k_{z}&=&k_{z}'e^{-\frac{\ell}{3}}, \label{Eq:kzScalingCNLF}
\\
\omega&=&\omega'e^{-\ell}, \label{Eq:omegaScalingCNLF}
\\
\psi&=&\psi' e^{\frac{4}{3}\ell},  \label{Eq:psiScalingCNLF}
\\
B&=&B', \label{Eq:BScalingCNLF}
\end{eqnarray}
the action can be written as
\begin{eqnarray}
S_{\psi'}=k_{F}\int\frac{d\omega'}{2\pi}\frac{dk_{r}'}{2\pi}\frac{dk_{z}'}{2\pi}\psi'^{\dag}(\omega',\mathbf{k}')
\left[-i\omega'+B'\left(k_{r}'^{3}-3k_{r}'k_{z}'^{2}\right)\sigma_{1}
+B'\left(k_{z}'^{3}-3k_{r}'k_{z}'^{2}\right)\sigma_{2}
\right]\psi'(\omega',\mathbf{k}'),
\end{eqnarray}
which has the same form as  the original action.

The action describing the four-fermion interactions between cubic NLFs is given by
\begin{eqnarray}
S_{\psi^{4}}
&=&\frac{1}{N}\sum_{i=0}^{3}\lambda_{i}k_{F}^{3}\int
\frac{d\omega_{1}}{2\pi}\frac{dk_{1r}}{2\pi}\frac{dk_{1z}}{2\pi}
\frac{d\omega_{2}}{2\pi}\frac{dk_{2r}}{2\pi}\frac{dk_{2z}}{2\pi}
\frac{d\omega_{3}}{2\pi}\frac{dk_{3r}}{2\pi}\frac{dk_{3z}}{2\pi}
\psi^{\dag}(\omega_{1},\mathbf{k}_{1})\sigma_{i}
\psi(\omega_{2},\mathbf{k}_{2})\psi^{\dag}(\omega_{3},\mathbf{k}_{3})\sigma_{i}\nonumber
\\
&&\times\psi(\omega_{1}-\omega_{2}+\omega_{3},
\mathbf{k}_{1}-\mathbf{k}_{2}+\mathbf{k}_{3}).
\end{eqnarray}
Including one-loop order corrections, the action becomes
\begin{eqnarray}
S_{\psi^{4}}
&=&\frac{1}{N}\sum_{i=0}^{3}\left(\lambda_{i}+\delta \lambda_{i}\right)k_{F}^{3}\int
\frac{d\omega_{1}}{2\pi}\frac{dk_{1r}}{2\pi}\frac{dk_{1z}}{2\pi}
\frac{d\omega_{2}}{2\pi}\frac{dk_{2r}}{2\pi}\frac{dk_{2z}}{2\pi}
\frac{d\omega_{3}}{2\pi}\frac{dk_{3r}}{2\pi}\frac{dk_{3z}}{2\pi}
\psi^{\dag}(\omega_{1},\mathbf{k}_{1})\sigma_{i}
\psi(\omega_{2},\mathbf{k}_{2})\psi^{\dag}(\omega_{3},\mathbf{k}_{3})\sigma_{i}\nonumber
\\
&&\times\psi(\omega_{1}-\omega_{2}+\omega_{3},
\mathbf{k}_{1}-\mathbf{k}_{2}+\mathbf{k}_{3}).
\end{eqnarray}
Utilizing the transformations shown in Eqs.~(\ref{Eq:krScalingCNLF})-(\ref{Eq:psiScalingCNLF}), and
\begin{eqnarray}
\lambda_{i}'&=&\left(\lambda_{i}+\delta\lambda_{i}\right)e^{\frac{1}{3}\ell}
\approx\lambda_{i}+\lambda_{i}\frac{1}{3}\ell+\delta\lambda_{i},
\label{Eq:lambdaScalingCNLF}
\end{eqnarray}
the action becomes
\begin{eqnarray}
S_{\psi'^{4}}
&=&\sum_{i=0}^{3}\lambda_{i}'k_{F}^{3}\int
\frac{d\omega_{1}'}{2\pi}\frac{dk_{1r}'}{2\pi}\frac{dk_{1z}'}{2\pi}
\frac{d\omega_{2}'}{2\pi}\frac{dk_{2r}'}{2\pi}\frac{dk_{2z}'}{2\pi}
\frac{d\omega_{3}'}{2\pi}\frac{dk_{3r}'}{2\pi}\frac{dk_{3z}'}{2\pi}
\psi'^{\dag}(\omega_{1}',\mathbf{k}_{1}')\sigma_{i}
\psi'(\omega_{2}',\mathbf{k}_{2}')\psi'^{\dag}(\omega_{3}',\mathbf{k}_{3}')\sigma_{i}\nonumber
\\
&&\times\psi'(\omega_{1}'-\omega_{2}'+\omega_{3}',
\mathbf{k}_{1}'-\mathbf{k}_{2}'+\mathbf{k}_{3}'),
\end{eqnarray}
which has the same form of the original action. According to  Eq.~(\ref{Eq:lambdaScalingCNLF}), the RG equation for $\lambda_{i}$ is given by
\begin{eqnarray}
\frac{d\lambda_{i}}{d\ell}=\frac{1}{3}\lambda_{i}+\frac{d\delta\lambda_{i}}{d\ell}.
\label{Eq:RGElambdaGeneralCNLF}
\end{eqnarray}
Substituting Eqs.~(\ref{Eq:deltalambda0CNLF})-(\ref{Eq:deltalambda3CNLF}) into Eq.~(\ref{Eq:RGElambdaGeneralCNLF}), the RG equations for $\lambda_{i}$ can be
written as

\begin{eqnarray}
\frac{d\lambda_{0}}{d\ell}&=&\frac{1}{3}\lambda_{0}+
\left(\lambda_{1}\lambda_{3}+\lambda_{2}\lambda_{3}\right)\frac{1}{N},
\\
\frac{d\lambda_{1}}{d\ell}&=&\frac{1}{3}\lambda_{1}+\left(-\lambda_{1}^{2}+\lambda_{0}\lambda_{1}+\lambda_{0}\lambda_{3}
+\lambda_{1}\lambda_{2}+\lambda_{1}\lambda_{3}-2\lambda_{2}\lambda_{3}\right)\frac{1}{N}+\lambda_{1}^{2},
\\
\frac{d\lambda_{2}}{d\ell}&=&\frac{1}{3}\lambda_{2}+\left(-\lambda_{2}^{2}+\lambda_{0}\lambda_{2}+\lambda_{0}\lambda_{3}+\lambda_{1}\lambda_{2}
-2\lambda_{1}\lambda_{3}+\lambda_{2}\lambda_{3}\right)\frac{1}{N}+\lambda_{2}^{2},
\\
\frac{d\lambda_{3}}{d\ell}&=&\frac{1}{3}\lambda_{3}+\left(-2\lambda_{3}^{2}+\lambda_{0}\lambda_{1}+\lambda_{0}\lambda_{2}+2\lambda_{0}\lambda_{3}
-2\lambda_{1}\lambda_{2}+2\lambda_{1}\lambda_{3}+2\lambda_{2}\lambda_{3}\right)\frac{1}{N}
+2\lambda_{3}^{2}.
\end{eqnarray}
The transformations
\begin{eqnarray}
\frac{k_{F}}{2\pi B\Lambda}\lambda_{i}\rightarrow\lambda_{i},
\end{eqnarray}
with $i=0, 1, 2, 3$ have been adopted.

\end{widetext}


\begin{thebibliography}{99}

\bibitem{Kotov12}
V. N. Kotov, B. Uchoa, V. M. Pereira, F. Guinea, and A. H. Castro
Neto, Electron-electron interactions in graphene: Current status and
perspectives, Rev. Mod. Phys. {\bf 84}, 1067 (2012).

\bibitem{Vafek14}
O. Vafek and A. Vishwanath, Dirac fermions in solids: From
high-T$_{c}$ cuprates and graphene to topological insulators and
Weyl semimetals, Annu. Rev. Condens. Matter Phys. {\bf 5}, 83
(2014).

\bibitem{Wehling14}
T. O. Wehling, A. M. Black-Schaffer, and A. V. Balatsky, Dirac
materials, Adv. Phys. {\bf 63}, 1 (2014).

\bibitem{Wan11}
X. Wan, A. M. Turner, A. Vishwanath, and S. Y. Savrasov, Topological
semimetal and Fermi-arc surface states in the electronic structure
of pyrochlore iridates, Phys. Rev. B {\bf 83}, 205101 (2011).

\bibitem{Weng16}
H. Weng, X. Dai, and Z. Fang, Topological semimetals predicted from
first-principles calculations, J. Phys.: Condens. Matter {\bf 28},
303001 (2016).

\bibitem{FangChen16}
C. Fang, H. Weng, X. Dai, and Z. Fang, Topological nodal line
semimetals, Chin. Phys. B {\bf 25}, 117106 (2016).

\bibitem{Yan17}
B. Yan and C. Felser, Topological materials: Weyl semimetals, Annu.
Rev. Condens. Matter Phys. {\bf 8}, 337 (2017).

\bibitem{Hasan17}
M. Z. Hasan, S.-Y. Xu, I. Belopolski, and S.-M. Huang, Discovery of
Weyl fermion semimetals and topological Fermi arc states, Annu. Rev.
Condens. Matter Phys. {\bf 8}, 289 (2017).

\bibitem{Armitage18}
N. P. Armitage, E. J. Mele, and A. Vishwanath, Weyl and Dirac
semimetals in three-dimensional solids, Rev. Mod. Phys. {\bf 90},
015001 (2018).

\bibitem{WangJian18}
H. Wang and J. Wang, Electron transport in Dirac and Weyl semimetals, Chin. Phys. B {\bf 27},
107402 (2018).

\bibitem{LvQianDing19}
B. Lv, T. Qian, and H. Ding, Angle-resolved photoemission spectroscopy and its
application to topological materials, Nat. Rev. Phys. {\bf 1}, 609 (2019).

\bibitem{Kruthoff17}
J. Kruthoff, J. de Boer, J. van Wezel, C. L. Kane, and R.-J. Slager, Topological classification
of crystalline insulators through band structure combinatorics, Phys. Rev. X {\bf 7}, 041069 (2017).

\bibitem{Tang19Wan}
F. Tang, H. C. Po, A. Vishwanath, and X. Wan, Comprehensive
search for topological materials using symmetry indicators, Nature
{\bf 566}, 486 (2019).

\bibitem{Zhang19FangChen}
T. Zhang, Y. Jiang, Z. Song, H. Huang, Y. He, Z. Fang, H. Weng, and
C. Fang, Catalogue of topological electronic materials, Nature {\bf
566}, 475 (2019).

\bibitem{VergnioryWangZ19}
M. G. Vergniory, L. Elcoro, C. Felser, N. Regnault, B. A. Bernevig,
and Z. Wang, A complete catalogue of high-quality topological
materials, Nature {\bf 566}, 480 (2019).

\bibitem{Skinner18}
B. Skinner and L. Fu, Large, nonsaturating thermopower in a quantizing
magnetic field, Sci. Adv. {\bf 4}, eaat2621 (2018).

\bibitem{Markov19}
M. Markov. S. E. Razaei, S. N. Sadeghi, K. Esfarjani, and M. Zebarjadi,
Thermoelectric properties of semimetals, Phys. Rev. Materials {\bf 3}, 095401 (2019).

\bibitem{Neupane14}
M. Neupane, S.-Y. Xu, R. Sankar, N. Alidoust, G. Bian, C. Liu, I.
Belopolski, T.-R. Chang, H.-T. Jeng, H. Lin, A. Bansil, F. Chou, and M.
Z. Hasan, Observation of a three-dimensional topological Dirac
semimetal phase in high-mobility Cd$_{3}$As$_{2}$, Nat. Commun. {\bf 5}, 3786 (2014).

\bibitem{Liu14}
Z. K. Liu, B. Zhou, Y. Zhang, Z. J. Wang, H. M. Weng, D.
Prabhakaran, S.-K. Mo, Z. X. Shen, Z. Fang, X. Dai, Z. Hussain, and
Y. L. Chen, Discovery of a three-dimensional topological Dirac
semimetal, Na$_{3}$Bi, Science {\bf 343}, 864 (2014).

\bibitem{Xu15}
S.-Y. Xu, I. Belopolski, N. Alidoust, M. Neupane, G. Bian, C. Zhang,
R. Sankar, G. Chang, Z. Yuan, C.-C. Lee, S.-M. Huang, H. Zheng, J.
Ma, D. S. Sanchez, B. Wang, A. Bansil, F. Chou, P. P. Shibayev, H.
Lin, S. Jia, and M. Z. Hasan, Discovery of a Weyl fermion semimetal
and topological Fermi arcs, Science {\bf 349}, 613 (2015).

\bibitem{Lv15}
B. Q. Lv, H. M. Weng, B. B. Fu, X. P. Wang, H. Miao, J. Ma, P.
Richard, X. C. Huang, L. X. Zhao, G. F. Chen, Z. Fang, X. Dai, T.
Qian, and H. Ding, Experimental discovery of Weyl semimetal TaAs,
Phys. Rev. X {\bf 5}, 031013 (2015).

\bibitem{Bian16}
G. Bian, T.-R. Chang, R. Sankar, S.-Y. Xu, H. Zheng, T. Neupert,
C.-K. Chiu, S.-M. Huang, G. Chang, I. Belopolski, D. S. Sanchez, M.
Neupane, N. Alidoust, C. Liu, B. Wang, C.-C. Lee, H.-T. Jeng, C.
Zhang, Z. Yuan, S. Jia, A. Bansil, F. Chou, H. Lin, and M. Z. Hasan,
Topological nodal-line fermions in spin-orbit metal PbTaSe$_{2}$,
Nat. Commun. {\bf 7}, 10556 (2016).

\bibitem{Neupane16}
M. Neupane, I. Belopolski, M. M. Hosen, D. S. Sanchez, R. Sankar, M.
Szlawska, S.-Y. Xu, K. Dimitri, N. Dhakal, P. Maldonado, P. M.
Oppeneer, D. Kaczorowski, F. Chou, M. Z. Hasan, and T. Durakiewicz,
Observation of topological nodal fermion semimetal phase in ZrSiS,
Phys. Rev. B {\bf 93}, 201104(R) (2016).

\bibitem{Schoop16}
L. M. Schoop, M. N. Ali, C. Strasser, A. Topp, A. Varykhalov, D.
Marchenko, V. Duppel, S. S. P. Parkin, B. V. Lotsch, and C. R. Ast,
Dirac cone protected by non-symmorphic symmetry and
three-dimensional Dirac line node in ZrSiS, Nat. Commun. {\bf 7},
11696 (2016).

\bibitem{Hosen17}
M. M. Hosen, K. Dimitri, I. Belopolski, P. Maldonado, R. Sankar, N. Dhakal,
G. Dhakal, T. Cole, P. M. Oppeneer, D. Kaczorowski, F. Chou, M. Z. Hasan,
T. Durakiewicz, and M. Neupane, Tunability of the topological nodal-line
semimetal phase in ZrSi$X$-type materials ($X$ = S, Se, Te), Phys. Rev. B
{\bf 95}, 161101(R) (2017).


\bibitem{Takane16}
D. Takane, Z. Wang, S. Souma, K. Nakayama, C. X. Trang, T. Sato, T.
Takahashi, and Y. Ando, Dirac-node arc in the topological line-node
semimetal HfSiS, Phys. Rev. B {\bf 94}, 121108(R) (2016).

\bibitem{Yi18}
C.-J. Yi, B. Q. Lv, Q. S. Wu. B.-B. Fu, X. Gao, M. Yang, X.-L. Peng,
M. Li, Y.-B. Huang, P. Richard, M. Shi, G. Li, O. V. Yazyev, Y.-G.
Shi, T. Qian, and H. Ding, Observation of a nodal chain with Dirac
surface states in TiB$_{2}$, Phys. Rev. B {\bf 97}, 201107(R) (2018).

\bibitem{Liu18}
Z. Liu, R. Luo, P. Guo, Q. Wang, S. Sun, C. Li, S. Thirupathaiah, A.
Fedorov, D. Shen, K. Liu, H. Lei, and S. Wang, Experimental
observation of Dirac nodal link in centrosymmetric semimetal
TiB$_{2}$, Phys. Rev. X {\bf 8}, 031044 (2018).

\bibitem{GiulianiBook}
G. F. Giuliani and G. Vignale, \emph{Quantum Theory of the Electron
Liquid} (Cambridge University Press, 2005).

\bibitem{Gonzalez99}
J. Gonz\'{a}lez, F. Guinea, and M. A. H. Vozmediano,
Marginal-Fermi-liquid behavior from two-dimensional Coulomb
interaction, Phys. Rev. B {\bf 59}, R2474(R) (1999).

\bibitem{Scheehy07}
D. E. Sheehy and J. Schmalian, Quantum critical scaling in graphene,
Phys. Rev. Lett. {\bf 99}, 226803 (2007).

\bibitem{Hofmann14}
J. Hofmann, E. Barnes, and S. Das Sarma, Why does graphene behave as
a weakly interacting system?, Phys. Rev. Lett. {\bf 113}, 105502
(2014).

\bibitem{WangLiu14}
J.-R. Wang and G.-Z. Liu, Influence of Coulomb interaction on the anisotropic Dirac cone in graphene,
Phys. Rev. B {\bf 89}, 195404 (2014).

\bibitem{Goswami11}
P. Goswami and S. Chakravarty, Quantum criticality between
topological and band insulators in 3+1 dimensions, Phys. Rev. Lett.
{\bf 107}, 196803 (2011).

\bibitem{Hosur12}
P. Hosur, S. A. Parameswaran, and A. Vishwanath,
Charge transport in Weyl semimetals, Phys. Rev. Lett. {\bf 108},
046602 (2012).

\bibitem{Throckmorton15}
R. E. Throckmorton, J. Hofmann, E. Barnes, and S. Das Sarma,
Many-body effects and ultraviolet renormalization in
three-dimensional Dirac materials, Phys. Rev. B {\bf 92}, 115101
(2015).

\bibitem{Moon13}
E.-G. Moon, C. Xu, Y. B. Kim, and L. Balents, Non-Fermi-liquid and
topological states with strong spin-orbit coupling, Phys. Rev. Lett.
{\bf 111}, 206401 (2013).

\bibitem{Herbut14}
I. F. Herbut and L. Janssen, Topological Mott insulator in
three-dimensional systems with quadratic band touching, Phys. Rev.
Lett. {\bf 113}, 106401 (2014).

\bibitem{YangNatPhys14}
B.-J. Yang, E.-G. Moon, H. Isobe, and N. Nagaosa, Quantum
criticality of topological phase transitions in three-dimensional
interacting electronic systems, Nat. Phys. {\bf 10}, 774 (2014).

\bibitem{Isobe16}
H. Isobe, B.-J. Yang, A. Chubukov, J. Schmalian, and N. Nagaosa,
Emergent non-Fermi-liquid at the quantum critical point of a
topological phase transition in two dimensions, Phys. Rev. Lett.
{\bf 116}, 076803 (2016).

\bibitem{WangLiuZhang17A}
J.-R. Wang, G.-Z. Liu, and C.-J. Zhang, Excitonic pairing and
insulating transition in two-dimensional semi-Dirac semimetals,
Phys. Rev. B {\bf 95}, 075129 (2017).

\bibitem{Lai15}
H.-H. Lai, Correlation effects in double-Weyl semimetals, Phys. Rev. B {\bf 91}, 235131 (2015).

\bibitem{Jian15}
S.-K. Jian and H. Yao, Correlated double-Weyl semimetals with
Coulomb interactions: Possible applications to HgCr$_2$Se$_4$ and
SrSi$_2$, Phys. Rev. B {\bf 92}, 045121 (2015).

\bibitem{WangLiuZhang17B}
J.-R. Wang, G.-Z. Liu, and C.-J. Zhang, Quantum phase transition
and unusual critical behavior in multi-Weyl semimetals, Phys. Rev.
B {\bf 96}, 165142 (2017).

\bibitem{ZhangShiXin17}
S.-X. Zhang, S.-K. Jian, and H. Yao, Correlated triple-Weyl
semimetals with Coulomb interactions, Phys. Rev. B {\bf 96},
241111(R) (2017).

\bibitem{WangLiuZhang18}
J.-R. Wang, G.-Z. Liu, and C.-J. Zhang, Breakdown of Fermi liquid
theory in topological multi-Weyl semimetals, Phys. Rev. B
{\bf 98}, 205113 (2018).

\bibitem{Huh16}
Y. Huh, E.-G. Moon, and Y. B. Kim, Long-range Coulomb interaction in
nodal-ring semimetals, Phys. Rev. B {\bf 93}, 035138 (2016).

\bibitem{WangYuXuan17}
Y. Wang and R. M. Nandkishore, Interplay between short-range correlated disorder and
Coulomb interaction in nodal-line semimetals, Phys. Rev. B {\bf 96}, 115130 (2017).

\bibitem{WangLiuZhang19}
J.-R. Wang, G.-Z. Liu, and C.-J. Zhang, Topological quantum critical point in a
triple-Weyl semimetal: Non-Fermi-liquid behavior and instabilities,
Phys. Rev. B {\bf 99}, 195119 (2019).

\bibitem{Han19}
S. Han, C. Lee, E.-G. Moon, and H. Min, Emergent anisotropic non-Fermi liquid
at a topological phase transition in three dimensions,
Phys. Rev. Lett. {\bf 122}, 187601 (2019).

\bibitem{Zhang18}
S.-X. Zhang, S.-K. Jian, and H. Yao, Quantum criticality preempted
by nematicity, arXiv:1809.10686.

\bibitem{Han18A}
S. Han and E.-G. Moon, Long-range Coulomb interaction effects on the
topological phase transitions between semimetals and insulators, Phys.
Rev. B {\bf 97}, 241101(R) (2018).

\bibitem{Han18B}
S. Han, G. Y. Cho, and E.-G. Moon, Quantum criticality with infinite
anisotropy in topological phase transitions between Dirac and Weyl semimetals,
Phys. Rev. B {\bf 98}, 085149 (2018).

\bibitem{Roy18Birefringent}
B. Roy, M. P. Kennett, K. Yang, and V. Juri\v{c}i\'{c}, From birefringent
electrons to a marginal or non-Fermi liquid of relativistic spin-$1/2$
fermions: An emergent superuniversality, Phys. Rev. Lett. {\bf 121}, 157602 (2018).

\bibitem{Uryszek19}
M. D. Uryszek, E. Christou, A. Jaefari, F. Kr\"{u}ger, and B. Uchoa, Quantum criticality of semi-Dirac fermions
in $2 + 1$ dimensions, Phys. Rev. B {\bf 100}, 155101 (2019).

\bibitem{Sur19}
S. Sur and B. Roy, Unifying interacting nodal semimetals: A new route to strong coupling, Phys.
Rev. Lett. {\bf 123}, 207601 (2019).

\bibitem{Herbut06}
I. F. Herbut, Interactions and phase transitions on
graphene's honeycomb lattice, Phys. Rev. Lett. {\bf 97}, 146401 (2006).

\bibitem{Herbut09}
I. F. Herbut, V. Juri\v{c}i\'{c}, and B. Roy, Theory of interacting electrons on the honeycomb lattice, Phys. Rev. B {\bf 79}, 085116 (2009).

\bibitem{Roy16}
B. Roy and S. Das Sarma, Quantum phases of interacting electrons in three-dimensional dirty Dirac semimetals, Phys. Rev. B {\bf 94}, 115137 (2016).

\bibitem{Maciejko14}
J. Maciejko and R. Nandkishore, Weyl semimetals with short-range interactions,
Phys. Rev. B {\bf 90}, 035126 (2014).

\bibitem{Roy17B}
B. Roy, P. Goswami, and V. Juri\v{c}i\'{c}, Interacting Weyl fermions: Phases, phase transitions, and global phase diagram, Phys. Rev. B {\bf 95}, 201102(R) (2017).

\bibitem{Roy18A}
B. Roy and M. S. Foster, Quantum multicriticality near the Dirac-semimetal to band-insulator critical point in two dimensions: A controlled ascent from one
dimension, Phys. Rev. X {\bf 8}, 011049 (2018).

\bibitem{WangJing18}
J. Wang, Role of four-fermion interaction and impurity in the states of two-dimensional semi-Dirac materials,
J. Phys.: Condens. Matter {\bf 30}, 125401 (2018).

\bibitem{Roy18B}
A. L. Szab\'{o}, R. Moessner, and B. Roy, Interacting spin-3/2 fermions in a Luttinger (semi)metal:
competing phases and their selection in the global phase diagram, arXiv:1811.12415.

\bibitem{Sur16}
S. Sur and R. Nandkishore, Instabilities of Weyl loop semimetals, New J. Phys. {\bf 18}, 115006 (2016).

\bibitem{Roy17A}
B. Roy, Interacting nodal-line semimetal: Proximity effect and spontaneous symmetry
breaking, Phys. Rev. B {\bf 96}, 041113(R) (2017).

\bibitem{Araujo18}
M. A. N. Ara\'{u}jo and L. Li, Broken-symmetry phases of interacting nested Weyl and Dirac loops,
Phys. Rev. B {\bf 98}, 155114 (2018).

\bibitem{Syzranov17}
S. V. Syzranov and B. Skinner, Electron transport in nodal-line semimetals, Phys.
Rev. B {\bf 96}, 161105(R) (2017).

\bibitem{LiXieGroup18}
C. Li, C. M. Wang, B. Wan, X. Wan, H.-Z. Lu, and X. C. Xie, Rules for phase shifts of
quantum oscillations in topological nodal-line semimetals, Phys. Rev. Lett. {\bf 120},
146602 (2018).

\bibitem{Chen19}
W. Chen, H.-Z. Lu, and O. Zilberberg, Weak localization and antilocalization in
nodal-line semimetals: Dimensionality and topological effects, Phys. Rev. Lett.
{\bf 122}, 196603 (2019).

\bibitem{Yu19}
Z.-M. Yu, W. Wu, X.-L. Sheng, Y. X. Zhao, and S. A. Yang, Quadratic and cubic nodal lines
stabilized by crystalline symmetry, Phys. Rev. B {\bf 99}, 121106(R) (2019).

\bibitem{LiLinHu17}
L. Li, S. Chesi, C. Yin, and S. Chen, $2\pi$-flux loop semimetals, Phys. Rev. B {\bf 96}, 081116(R) (2017).

\bibitem{Shankar94}
R. Shankar, Renormalization-group approach to interacting fermions, Rev. Mod. Phys. {\bf 66},
129 (1994).

\bibitem{Nandkishore12}
R. Nandkishore, L. S. Levitov, and A. V. Chubukov, Chiral superconductivity from repulsive
interactions in doped graphene, Nat. Phys. {\bf 8}, 158 (2012).

\bibitem{Metlitski15}
M. A. Metlitski, D. F. Mross, S. Sachdev, and T. Senthil, Cooper pairing in non-Fermi liquids,
Phys. Rev. B {\bf 91}, 115111 (2015).

\bibitem{Vafek10}
O. Vafek and K. Yang, Many-body instability of Coulomb interacting bilayer graphene: Renormalization group approach, Phys. Rev. B {\bf 81}, 041401(R) (2010).

\bibitem{Zhang10}
F. Zhang, H. Min, M. Polini, and A. H. MacDonald, Spontaneous inversion symmetry breaking in graphene bilayers, Phys. Rev. B {\bf 81}, 041402(R) (2010).

\bibitem{Isobe19}
H. Isobe and L. Fu, Supermetal, Phys. Rev. Research {\bf 1}, 033206 (2019).

\bibitem{Volkov18}
P. A. Volkov and S. Moroz, Coulomb-induced instabilities of nodal surfaces, Phys. Rev. B {\bf 98}, 241107(R) (2018).



























\end{thebibliography}
\end{document}